\documentclass[preprint,aps, showpacs]{revtex4}
\usepackage{graphicx}
\usepackage{amsmath}
\begin{document}
\draft
\title{ Transformation design approach to a fermionic spacetime cage}
\author{De-Hone Lin \thanks{%
e-mail: dhlin@mail.nsysu.edu.tw}}
\address{Department of Physics, National Sun Yat-sen University, Kaohsiung, Taiwan}
\date{\today }
\begin{abstract}
This paper is concerned with the application of a spacetime structure 
which emerges from the transformation design of spin-1/2 fermions to a
three-dimensional quantum system. There are three components. First, the
main part of this paper presents the constraint conditions which build the
relation of a spacetime structure and a form invariance solution to the
covariant Dirac equation. The second is to devise a spacetime cage for
fermions with chosen constraints. The third part discusses 
the feasibility of the cage with an experiment.
\end{abstract}
\pacs{04.40.-b, 03.65.Pm, 11.80.Et, 03.75.-b\\
Keywords: Dirac Equation; Constraint Conditions; Spacetime Structure; Form Invariance Solution\\
Email: dhlin@mail.nsysu.edu.tw}
\maketitle
\tolerance=10000
\section{INTRODUCTION}

Searching for a form invariance solution to the Maxwell equations is the
central issue in making an invisibility cloak for electromagnetic waves with
the transformation design method \cite{1,2}. Recently, it was shown that the
form invariance solution for the Dirac equation establishes the relation
between the quantum rule of two-dimensional (2D) fermions in a central force
field and a spacetime structure through constraint conditions \cite{3}.

In this paper, we present the conditions to obtain a 3D quantum system with
a form invariance spinor. As an application, a spacetime cage for fermions
is coined. The paper is arranged as follows: In Section II, the constraints
which determine the spacetime and the 3D form invariance solution for
fermions in the central force field are presented. We devote Section III to
constructing the spacetime cage which traps fermion states accompanying an
invisible region by the chosen constraints and thereby the spacetime
structure emerged. The physical approach to the trapping effect by way of an
experiment is outlined. Our conclusion and several notes are summarized in
Section IV. Finally, the appendices add some detailed mathematical support
to the discussions involving the length computations in Section II.

\section{Constraints and the form invariance Dirac spinor}

We discuss the conditions to achieve a form invariance wave function for
fermions in this section. The evolution of spin-1/2 fermions with mass $M$
in curved spacetime is depicted by the spinor $\Phi $ which satisfies the
covariant Dirac equation (e.g. \cite{4})%
\begin{equation}
\left\{ \tilde{\gamma}^{\mu }\left( \partial _{\mu }-\Gamma _{\mu }\right)
+M\right\} \Phi =0,  \label{01}
\end{equation}%
where $\partial _{\mu }=\partial /\partial x^{\mu \text{ }}$with $x^{\mu
\text{ }}=(x^{1},x^{2},x^{3},x^{4})=(x,y,z,ict)$ in our convention, the
symbol $\tilde{\gamma}^{\mu }$ labels the spin matrices in curved space, and
$\Gamma _{\mu }$ is the spin connection. The fundamental connection between
spin and spacetime is through the anticommutation relation%
\begin{equation}
\tilde{\gamma}^{\mu }\tilde{\gamma}^{\nu }+\tilde{\gamma}^{\nu }\tilde{\gamma%
}^{\mu }=2g^{\mu \nu },  \label{02}
\end{equation}%
in which $g^{\mu \nu }$ is the inverse of the metric of the line element%
\begin{equation}
ds^{2}=\sum_{\mu ,\nu =1}^{4}g_{\mu \nu }dx^{\mu }dx^{\nu },  \label{03}
\end{equation}%
satisfying $\sum_{\lambda }g_{\mu \lambda }g^{\lambda \nu }=\delta _{\mu
}^{\;\nu }$. Throughout this paper, the natural units $\hbar =c=1$ are
adopted if not explicitly stated otherwise. In the following discussions, we
shall restrict ourselves to spacetime described by the line element%
\begin{equation}
ds^{2}=f_{1}^{2}(dx^{1})^{2}+f_{2}^{2}(dx^{2})^{2}+f_{3}^{2}(dx^{3})^{2}+f_{4}^{2}(dx^{4})^{2}.
\label{04}
\end{equation}%
So we have the metric of spacetime%
\begin{equation}
g_{\mu \nu }={\rm diag}(g_{11},g_{22},g_{33},g_{44})={\rm diag}%
(f_{1}^{2},f_{2}^{2},f_{3}^{2},f_{4}^{2}),  \label{05}
\end{equation}%
where $f_{\mu }$ are the arbitrary functions of the variables $%
x^{1},x^{2},x^{3},\ $and $x^{4}$. Based on the orthogonal assumption, the
spin connection has the representation \cite{3,5}%
\begin{equation}
\Gamma _{\mu }=\frac{1}{4}\left[ \left( \partial _{\mu }g_{\mu \mu }\right)
g^{\mu \mu }-(\partial _{\lambda }g_{\mu \mu })\tilde{\gamma}^{\mu }\tilde{%
\gamma}^{\lambda }\right] .  \label{06}
\end{equation}%
There is no summation over $\mu $ here. In the light of (\ref{06}), the
covariant Dirac equation can be expressed in terms of
\begin{equation}
\left\{ \sum_{i=1}^{4}\frac{\gamma ^{i}}{f_{i}}\left( \partial
_{i}-A_{i}\right) +M\right\} \tilde{\Psi}=0,  \label{07}
\end{equation}%
where the spinor $\tilde{\Psi}$ is connected to the old one by $\tilde{\Psi}%
=(f_{1}f_{2}f_{3}f_{4})\Phi =\sqrt{H}\Phi $ in which $H$ denotes the
determinant of the metric $g_{\mu \nu }$, the effective four vector%
\begin{equation}
A_{i}=\frac{1}{2}\partial _{i}\left[ \ln \left( f_{1}f_{2}f_{3}f_{4}\right)
+\ln f_{i}\right] ,  \label{08}
\end{equation}%
and $\gamma ^{i}$ are the spin matrices in the flat spacetime of special
relativity which satisfy the anticommutation relation
\begin{equation}
\gamma ^{i}\gamma ^{j}+\gamma ^{j}\gamma ^{i}=2\delta ^{ij}.  \label{09}
\end{equation}%
For a static spacetime, the metric is only the spacial function, i.e., $%
g_{\mu \mu }=f_{\mu }^{2}({\bf x})$. There exists a solution to the steady
state for $\tilde{\Psi}$ which is that the Dirac spinor can be decomposed as
$\tilde{\Psi}({\bf x,}t)=\Psi ({\bf x})\exp \{-iEt\}$. Eq. (\ref{07}) then
reduces to%
\begin{equation}
\left\{ \left[ \sum_{i=1}^{3}\frac{\gamma ^{i}}{f_{i}}\left( \partial
_{i}-A_{i}\right) \right] -\frac{\gamma ^{4}}{f_{4}}E+M\right\} \Psi ({\bf x}%
)=0.  \label{010}
\end{equation}%
We are now in a position to seek a form invariance solution in the line
below with the formulation. For this, we assume the metric $g_{\mu \mu
}=f_{\mu }^{2}(R(r))$ with $R(r)$ being arbitrary radial function, the four
component spinor is taken as $\Psi ({\bf x})=(\Phi _{1},\Phi _{2},\Phi
_{3},\Phi _{4})^{T}$, and the spin matrices conforming to (\ref{09}) are
chosen to be%
\begin{equation}
\gamma ^{k}=\left(
\begin{array}{ll}
0 & -i\sigma _{k} \\
i\sigma _{k} & 0%
\end{array}%
\right) \text{, }k=1,2,3\text{, and }\gamma ^{4}=\left(
\begin{array}{ll}
I & 0 \\
0 & -I%
\end{array}%
\right) .  \label{011}
\end{equation}%
Substituting the representation into (\ref{010}), the Dirac equation has the
decomposition%
\[
-i\left\{ \frac{1}{f_{1}}\partial _{x}\Phi _{4}-\frac{xR^{\prime }(r)}{%
2rf_{1}}\left[ \frac{d}{dR}\left( \ln f_{1}f_{2}f_{3}f_{4}+\ln f_{1}\right) %
\right] \Phi _{4}\right\}
\]%
\[
-\left\{ \frac{1}{f_{2}}\partial _{y}\Phi _{4}-\frac{yR^{\prime }(r)}{2rf_{2}%
}\left[ \frac{d}{dR}\left( \ln f_{1}f_{2}f_{3}f_{4}+\ln f_{2}\right) \right]
\Phi _{4}\right\}
\]%
\[
-i\left\{ \frac{1}{f_{3}}\partial _{z}\Phi _{3}-\frac{zR^{\prime }(r)}{%
2rf_{3}}\left[ \frac{d}{dR}\left( \ln f_{1}f_{2}f_{3}f_{4}+\ln f_{3}\right) %
\right] \Phi _{3}\right\}
\]%
\begin{equation}
-(E/f_{4}-M)\Phi _{1}=0,  \label{012}
\end{equation}%
\[
-i\left\{ \frac{1}{f_{1}}\partial _{x}\Phi _{3}-\frac{xR^{\prime }(r)}{%
2rf_{1}}\left[ \frac{d}{dR}\left( \ln f_{1}f_{2}f_{3}f_{4}+\ln f_{1}\right) %
\right] \Phi _{3}\right\}
\]%
\[
+\left\{ \frac{1}{f_{2}}\partial _{y}\Phi _{3}-\frac{yR^{\prime }(r)}{2rf_{2}%
}\left[ \frac{d}{dR}\left( \ln f_{1}f_{2}f_{3}f_{4}+\ln f_{2}\right) \right]
\Phi _{3}\right\}
\]%
\[
+i\left\{ \frac{1}{f_{3}}\partial _{z}\Phi _{4}-\frac{zR^{\prime }(r)}{%
2rf_{3}}\left[ \frac{d}{dR}\left( \ln f_{1}f_{2}f_{3}f_{4}+\ln f_{3}\right) %
\right] \Phi _{4}\right\}
\]%
\begin{equation}
-(E/f_{4}-M)\Phi _{2}=0,  \label{013}
\end{equation}%
\[
i\left\{ \frac{1}{f_{1}}\partial _{x}\Phi _{2}-\frac{xR^{\prime }(r)}{2rf_{1}%
}\left[ \frac{d}{dg}\left( \ln f_{1}f_{2}f_{3}f_{4}+\ln f_{1}\right) \right]
\Phi _{2}\right\}
\]%
\[
+\left\{ \frac{1}{f_{2}}\partial _{y}\Phi _{2}-\frac{yR^{\prime }(r)}{2rf_{2}%
}\left[ \frac{d}{dR}\left( \ln f_{1}f_{2}f_{3}f_{4}+\ln f_{2}\right) \right]
\Phi _{2}\right\}
\]%
\[
+i\left\{ \frac{1}{f_{3}}\partial _{z}\Phi _{1}-\frac{zR^{\prime }(r)}{%
2rf_{3}}\left[ \frac{d}{dR}\left( \ln f_{1}f_{2}f_{3}f_{4}+\ln f_{3}\right) %
\right] \Phi _{1}\right\}
\]%
\begin{equation}
+(E/f_{4}+M)\Phi _{3}=0,  \label{014}
\end{equation}%
and%
\[
i\left\{ \frac{1}{f_{1}}\partial _{x}\Phi _{1}-\frac{xR^{\prime }(r)}{2rf_{1}%
}\left[ \frac{d}{dR}\left( \ln f_{1}f_{2}f_{3}f_{4}+\ln f_{1}\right) \right]
\Phi _{1}\right\}
\]%
\[
-\left\{ \frac{1}{f_{2}}\partial _{y}\Phi _{1}-\frac{yR^{\prime }(r)}{2rf_{2}%
}\left[ \frac{d}{dR}\left( \ln f_{1}f_{2}f_{3}f_{4}+\ln f_{2}\right) \right]
\Phi _{1}\right\}
\]%
\[
-i\left\{ \frac{1}{f_{3}}\partial _{z}\Phi _{2}-\frac{zR^{\prime }(r)}{%
2rf_{3}}\left[ \frac{d}{dR}\left( \ln f_{1}f_{2}f_{3}f_{4}+\ln f_{3}\right) %
\right] \Phi _{2}\right\}
\]%
\begin{equation}
+(E/f_{4}+M)\Phi _{4}=0,  \label{015}
\end{equation}%
where $R^{\prime }(r)=dR(r)/dr$. A further reduction is made by assuming
that the angular part of the spinor $\Psi $ can be separated by the spin
spherical harmonics which are classified by the total angular momentum $j$,
and then $\Psi $ can be represented by%
\begin{equation}
\left\{
\begin{array}{l}
\Phi _{1}=\sqrt{\frac{l+m+1/2}{2l+1}}F(R(r))Y_{l,m-1/2} \\
\Phi _{2}=-\sqrt{\frac{l-m+1/2}{2l+1}}F(R(r))Y_{l,m+1/2} \\
\Phi _{3}=-i\sqrt{\frac{l-m+3/2}{2l+3}}G(R(r))Y_{l+1,m-1/2} \\
\Phi _{4}=-i\sqrt{\frac{l+m+3/2}{2l+3}}G(R(r))Y_{l+1,m+1/2}%
\end{array}%
\right. ,\text{ for }j=l+\frac{1}{2},\text{ }l=0,1,\cdots ,  \label{016}
\end{equation}%
and%
\begin{equation}
\left\{
\begin{array}{l}
\Phi _{1}=\sqrt{\frac{l-m+1/2}{2l+1}}F(R(r))Y_{l,m-1/2} \\
\Phi _{2}=\sqrt{\frac{l+m+1/2}{2l+1}}F(R(r))Y_{l,m+1/2} \\
\Phi _{3}=-i\sqrt{\frac{l+m-1/2}{2l-1}}G(R(r))Y_{l-1,m-1/2} \\
\Phi _{4}=i\sqrt{\frac{l-m-1/2}{2l-1}}G(R(r))Y_{l-1,m+1/2}%
\end{array}%
\right. ,\text{ for }j=l-1/2,\text{ }l=1,2,\cdots ,  \label{017}
\end{equation}%
where $m=m_{l}+1/2$ with $-l\leq m_{l}\leq l$. It is easy to show that the
chosen angular parts of the spinors are normalized to unity. After a tedious
calculation for the solution (\ref{016}), one verifies that equation (\ref%
{012}) is equivalent to the following four independent equalities (see
Appendix A):%
\begin{equation}
\left( -\frac{1}{f_{1}}+\frac{1}{f_{2}}\right) \left[ \frac{dG}{dr}-\left(
\frac{l+1}{r}\right) G\right] +\left[ -\frac{(\Delta _{1})}{f_{1}}+\frac{%
(\Delta _{2})}{f_{2}}\right] \frac{R^{\prime }}{2}G=0,  \label{018}
\end{equation}

\begin{equation}
\left( \frac{1}{f_{1}}-\frac{1}{f_{2}}\right) \left[ \frac{dG}{dr}+\left(
\frac{l+2}{r}\right) G\right] +\left[ \frac{(\Delta _{1})}{f_{1}}-\frac{%
(\Delta _{2})}{f_{2}}\right] \frac{R^{\prime }}{2}G=0,  \label{019}
\end{equation}%
\begin{equation}
\left( \frac{1}{2f_{1}}+\frac{1}{2f_{2}}-\frac{1}{f_{3}}\right) \left[ \frac{%
dG}{dr}-\left( \frac{l+1}{r}\right) G\right] +\left[ \frac{(\Delta _{1})}{%
2f_{1}}+\frac{(\Delta _{2})}{2f_{2}}+\frac{(\Delta _{3})}{f_{3}}\right]
\frac{R^{\prime }}{2}G=0,  \label{020}
\end{equation}%
and%
\[
-\left( \frac{l+m+3/2}{2(2l+3)}\right) \left\{ \left( \frac{1}{f_{1}}+\frac{1%
}{f_{2}}\right) \left[ \frac{dG}{dr}+\left( \frac{l+2}{r}\right) G\right] +%
\left[ \frac{(\Delta _{1})}{f_{1}}+\frac{(\Delta _{2})}{f_{2}}\right] \frac{%
R^{\prime }}{2}G\right\}
\]%
\begin{equation}
+\left( \frac{l-m+3/2}{2l+3}\right) \left\{ -\frac{1}{f_{3}}\left[ \frac{dG}{%
dr}+\left( \frac{l+2}{r}\right) G\right] +\frac{(\Delta _{3})}{f_{3}}\frac{%
R^{\prime }}{2}G\right\} +\left( -\frac{E}{f_{4}}+M\right) F=0.  \label{021}
\end{equation}%
For simplicity of presentation, here we have used $\Delta _{i}$ to denote%
\begin{equation}
\left\{
\begin{array}{c}
\Delta _{1}=\frac{d}{dR}\left( \ln f_{1}f_{2}f_{3}f_{4}+\ln f_{1}\right) ,
\\
\Delta _{2}=\frac{d}{dR}\left( \ln f_{1}f_{2}f_{3}f_{4}+\ln f_{2}\right) ,
\\
\Delta _{3}=\frac{d}{dR}\left( \ln f_{1}f_{2}f_{3}f_{4}+\ln f_{3}\right) .%
\end{array}%
\right.  \label{022}
\end{equation}%
Substituting (\ref{016}) into (\ref{013}), one obtains the four equations
equivalent to (\ref{013}) (see Appendix A):%
\begin{equation}
\left( \frac{1}{f_{1}}-\frac{1}{f_{2}}\right) \left[ \frac{dG}{dr}-\left(
\frac{l+1}{r}\right) G\right] +\left[ \frac{(\Delta _{1})}{f_{1}}-\frac{%
(\Delta _{2})}{f_{2}}\right] \frac{R^{\prime }}{2}G=0,  \label{023}
\end{equation}

\begin{equation}
\left( -\frac{1}{f_{1}}+\frac{1}{f_{2}}\right) \left[ \frac{dG}{dr}+\left(
\frac{l+2}{r}\right) G\right] +\left[ -\frac{(\Delta _{1})}{f_{1}}+\frac{%
(\Delta _{2})}{f_{2}}\right] \frac{R^{\prime }}{2}G=0,  \label{024}
\end{equation}%
\begin{equation}
\left( \frac{1}{2f_{1}}+\frac{1}{2f_{2}}-\frac{1}{f_{3}}\right) \left[ \frac{%
dG}{dr}-\left( \frac{l+1}{r}\right) G\right] +\left[ \frac{(\Delta _{1})}{%
2f_{1}}+\frac{(\Delta _{2})}{2f_{2}}+\frac{(\Delta _{3})}{f_{3}}\right]
\frac{R^{\prime }}{2}G=0,  \label{025}
\end{equation}%
and%
\[
\left( \frac{l-m+3/2}{2(2l+3)}\right) \left\{ \left( \frac{1}{f_{1}}+\frac{1%
}{f_{2}}\right) \left[ \frac{dG}{dr}+\left( \frac{l+2}{r}\right) G\right] +%
\left[ \frac{(\Delta _{1})}{f_{1}}+\frac{(\Delta _{2})}{f_{2}}\right] \frac{%
R^{\prime }}{2}G\right\}
\]%
\begin{equation}
+\left( \frac{l+m+3/2}{2l+3}\right) \left\{ \frac{1}{f_{3}}\left[ \frac{dG}{%
dr}+\left( \frac{l+2}{r}\right) G\right] -\frac{(\Delta _{3})}{f_{3}}\frac{%
R^{\prime }}{2}G\right\} +\left( \frac{E}{f_{4}}-M\right) F=0.  \label{026}
\end{equation}%
Note that Eqs. (\ref{018})-(\ref{020}) are the same as (\ref{023})-(\ref{025}%
). If we subtract (\ref{021}) from (\ref{026}), it follows that%
\[
\left\{ \left( \frac{1}{2f_{1}}+\frac{1}{2f_{2}}+\frac{1}{f_{3}}\right) %
\left[ \frac{dG}{dr}+\left( \frac{l+2}{r}\right) G\right] +\left[ \frac{%
(\Delta _{1})}{2f_{1}}+\frac{(\Delta _{2})}{2f_{2}}-\frac{(\Delta _{3})}{%
f_{3}}\right] \frac{R^{\prime }}{2}G\right\}
\]%
\begin{equation}
+2\left( \frac{E}{f_{4}}-M\right) F=0.  \label{027}
\end{equation}%
Eliminating the final term $(\Delta _{3})R^{\prime }G/2f_{3}$ in the second
middle bracket from Eq. (\ref{025}), we get%
\[
\left( \frac{1}{2f_{1}}+\frac{1}{2f_{2}}\right) \frac{dG}{dr}+\left[ \left(
\frac{1}{2f_{1}}+\frac{1}{2f_{2}}\right) \frac{1}{2r}+\left( \frac{(\Delta
_{1})}{2f_{1}}+\frac{(\Delta _{2})}{2f_{2}}\right) \frac{R^{\prime }}{2}+%
\frac{1}{f_{3}}\left( \frac{l+3/2}{r}\right) \right] G
\]%
\begin{equation}
+\left( \frac{E}{f_{4}}-M\right) F=0.  \label{028}
\end{equation}%
Changing the variable of differentiation by letting $d/dr=R^{\prime }d/dR$,
it turns into
\[
\frac{dG}{dR}+\frac{1}{R}\left[ \frac{R}{2rR^{\prime }}+\frac{R}{2}\frac{%
f_{1}(\Delta _{2})+f_{2}(\Delta _{1})}{f_{1}+f_{2}}\right] G+\left[ \frac{R}{%
rf_{3}R^{\prime }}\left( \frac{2f_{1}f_{2}}{f_{1}+f_{2}}\right) \right]
\left( \frac{l+3/2}{R}\right) G
\]%
\begin{equation}
+\frac{1}{R^{\prime }}\left( \frac{2f_{1}f_{2}}{f_{1}+f_{2}}\right) \left(
\frac{E}{f_{4}}-M\right) F=0.  \label{029}
\end{equation}%
Substituting (\ref{016}) into (\ref{014}) and (\ref{015}), and using the
same argument as above gives%
\[
\frac{dF}{dR}+\frac{1}{R}\left[ \frac{R}{2rR^{\prime }}+\frac{R}{2}\frac{%
f_{1}(\Delta _{2})+f_{2}(\Delta _{1})}{f_{1}+f_{2}}\right] F-\left[ \frac{R}{%
rf_{3}R^{\prime }}\left( \frac{2f_{1}f_{2}}{f_{1}+f_{2}}\right) \right]
\left( \frac{l+1/2}{R}\right) F
\]%
\begin{equation}
-\frac{1}{R^{\prime }}\left( \frac{2f_{1}f_{2}}{f_{1}+f_{2}}\right) \left(
\frac{E}{f_{4}}+M\right) G=0.  \label{040}
\end{equation}%
Combining with the chosen constraint conditions, Eqs. (\ref{029}) and (\ref%
{040}) will lead to form invariance solutions for fermions. Before doing
this, let's turn our attention to the equations that satisfy the trial
solution (\ref{017}). Substituting (\ref{017}) for (\ref{012}) gives the
four equivalent equations (see Appendix B)%
\begin{equation}
\left( \frac{1}{f_{1}}-\frac{1}{f_{2}}\right) \left[ \frac{dG}{dr}-\left(
\frac{l-1}{r}\right) G\right] +\left[ \frac{(\Delta _{1})}{f_{1}}-\frac{%
(\Delta _{2})}{f_{2}}\right] \frac{R^{\prime }}{2}G=0,  \label{041}
\end{equation}

\begin{equation}
\left( -\frac{1}{f_{1}}+\frac{1}{f_{2}}\right) \left[ \frac{dG}{dr}+\frac{l}{%
r}G\right] +\left[ -\frac{(\Delta _{1})}{f_{1}}+\frac{(\Delta _{2})}{f_{2}}%
\right] \frac{R^{\prime }}{2}G=0,  \label{042}
\end{equation}%
\begin{equation}
\left( \frac{1}{2f_{1}}+\frac{1}{2f_{2}}-\frac{1}{f_{3}}\right) \left[ \frac{%
dG}{dr}+\frac{l}{r}G\right] +\left[ \frac{(\Delta _{1})}{2f_{1}}+\frac{%
(\Delta _{2})}{2f_{2}}+\frac{(\Delta _{3})}{f_{3}}\right] \frac{R^{\prime }}{%
2}G=0,  \label{043}
\end{equation}%
and%
\[
-\left( \frac{l-m-1/2}{2(2l-1)}\right) \left\{ \left( \frac{1}{f_{1}}+\frac{1%
}{f_{2}}\right) \left[ \frac{dG}{dr}-\left( \frac{l-1}{r}\right) G\right] +%
\left[ \frac{(\Delta _{1})}{f_{1}}+\frac{(\Delta _{2})}{f_{2}}\right] \frac{%
R^{\prime }}{2}G\right\}
\]%
\begin{equation}
-\left( \frac{l+m-1/2}{2l-1}\right) \left\{ \frac{1}{f_{3}}\left[ \frac{dG}{%
dr}-\left( \frac{l-1}{r}\right) G\right] -\frac{(\Delta _{3})}{f_{3}}\frac{%
R^{\prime }}{2}G\right\} -\left( \frac{E}{f_{4}}+M\right) F=0.  \label{044}
\end{equation}

Substituting (\ref{017}) for (\ref{013}) results in the equalities (see
Appendix B)%
\begin{equation}
\left( \frac{1}{f_{1}}-\frac{1}{f_{2}}\right) \left[ \frac{dG}{dr}-\left(
\frac{l-1}{r}\right) G\right] +\left[ \frac{(\Delta _{1})}{f_{1}}-\frac{%
(\Delta _{2})}{f_{2}}\right] \frac{R^{\prime }}{2}G=0,  \label{045}
\end{equation}

\begin{equation}
\left( -\frac{1}{f_{1}}+\frac{1}{f_{2}}\right) \left[ \frac{dG}{dr}+\frac{l}{%
r}G\right] +\left[ -\frac{(\Delta _{1})}{f_{1}}+\frac{(\Delta _{2})}{f_{2}}%
\right] \frac{R^{\prime }}{2}G=0,  \label{046}
\end{equation}%
\begin{equation}
\left( \frac{1}{2f_{1}}+\frac{1}{2f_{2}}-\frac{1}{f_{3}}\right) \left[ \frac{%
dG}{dr}+\frac{l}{r}G\right] +\left[ \frac{(\Delta _{1})}{2f_{1}}+\frac{%
(\Delta _{2})}{2f_{2}}+\frac{(\Delta _{3})}{f_{3}}\right] \frac{R^{\prime }}{%
2}G=0,  \label{047}
\end{equation}%
and%
\[
-\left( \frac{l+m-1/2}{2(2l-1)}\right) \left\{ \left( \frac{1}{f_{1}}+\frac{1%
}{f_{2}}\right) \left[ \frac{dG}{dr}-\left( \frac{l-1}{r}\right) G\right] +%
\left[ \frac{(\Delta _{1})}{f_{1}}+\frac{(\Delta _{2})}{f_{2}}\right] \frac{%
R^{\prime }}{2}G\right\}
\]%
\begin{equation}
-\left( \frac{l-m-1/2}{2l-1}\right) \left\{ \frac{1}{f_{3}}\left[ \frac{dG}{%
dr}-\left( \frac{l-1}{r}\right) G\right] -\frac{(\Delta _{3})}{f_{3}}\frac{%
R^{\prime }}{2}G\right\} -\left( \frac{E}{f_{4}}-M\right) F=0.  \label{048}
\end{equation}%
Again, we see the same structure among (\ref{041})-(\ref{043}) and (\ref{045}%
)-(\ref{047}). Adding (\ref{044}) and (\ref{048}), one gets the equation%
\[
\left\{ \left( \frac{1}{2f_{1}}+\frac{1}{2f_{2}}+\frac{1}{f_{3}}\right) %
\left[ \frac{dG}{dr}-\left( \frac{l-1}{r}\right) G\right] +\left[ \frac{%
(\Delta _{1})}{2f_{1}}+\frac{(\Delta _{2})}{2f_{2}}-\frac{(\Delta _{3})}{%
f_{3}}\right] \frac{R^{\prime }}{2}G\right\}
\]%
\begin{equation}
+2\left( \frac{E}{f_{4}}-M\right) F=0.  \label{049}
\end{equation}%
Solving $(\Delta _{3})R^{\prime }G/2f_{3}$ from (\ref{047}), and
substituting its representation into the corresponding term in (\ref{049}),
it yields%
\[
\left( \frac{1}{2f_{1}}+\frac{1}{2f_{2}}\right) \frac{dG}{dr}+\left[ \left(
\frac{1}{2f_{1}}+\frac{1}{2f_{2}}\right) \frac{1}{2r}+\left( \frac{(\Delta
_{1})}{2f_{1}}+\frac{(\Delta _{2})}{2f_{2}}\right) \frac{R^{\prime }}{2}-%
\frac{1}{f_{3}}\left( \frac{l-1/2}{r}\right) \right] G
\]%
\begin{equation}
+\left( \frac{E}{f_{4}}-M\right) F=0.  \label{050}
\end{equation}%
It follows from the change of variable $dG/dr=(dG/dR)(dR/dr)$ that
\[
\frac{dG}{dR}+\frac{1}{R}\left[ \frac{R}{2rR^{\prime }}+\frac{R}{2}\frac{%
f_{1}(\Delta _{2})+f_{2}(\Delta _{1})}{f_{1}+f_{2}}\right] G-\left[ \frac{R}{%
rf_{3}R^{\prime }}\left( \frac{2f_{1}f_{2}}{f_{1}+f_{2}}\right) \right]
\left( \frac{l-1/2}{R}\right) G
\]%
\begin{equation}
+\frac{1}{R^{\prime }}\left( \frac{2f_{1}f_{2}}{f_{1}+f_{2}}\right) \left(
\frac{E}{f_{4}}-M\right) F=0.  \label{051}
\end{equation}%
Substituting (\ref{017}) into (\ref{014}) and (\ref{015}), an argument
similar to the one used above leads to
\[
\frac{dF}{dR}+\frac{1}{R}\left[ \frac{R}{2rR^{\prime }}+\frac{R}{2}\frac{%
f_{1}(\Delta _{2})+f_{2}(\Delta _{1})}{f_{1}+f_{2}}\right] F+\left[ \frac{R}{%
rf_{3}R^{\prime }}\left( \frac{2f_{1}f_{2}}{f_{1}+f_{2}}\right) \right]
\left( \frac{l+1/2}{R}\right) F
\]%
\begin{equation}
-\frac{1}{R^{\prime }}\left( \frac{2f_{1}f_{2}}{f_{1}+f_{2}}\right) \left(
\frac{E}{f_{4}}+M\right) G=0.  \label{062}
\end{equation}%
Eqs. (\ref{051}) and (\ref{062}) are an alternative set of equations to
obtain the form invariance solution.

It is time to choose the constraints for the form invariance. By comparing
Eqs. (\ref{029}), (\ref{040}), (\ref{051}), and (\ref{062}), we choose
\begin{equation}
\frac{R}{rR^{\prime }}+\frac{f_{1}(\Delta _{2})+f_{2}(\Delta _{1})}{%
f_{1}+f_{2}}R=1,  \label{063}
\end{equation}%
\begin{equation}
\frac{R}{rf_{3}R^{\prime }}\left( \frac{2f_{1}f_{2}}{f_{1}+f_{2}}\right) =1,
\label{064}
\end{equation}%
\begin{equation}
\frac{1}{R^{\prime }}\left( \frac{2f_{1}f_{2}}{f_{1}+f_{2}}\right) =1,
\label{065}
\end{equation}%
\begin{equation}
\frac{1}{f_{4}}=\text{any function of }R(r),  \label{066}
\end{equation}%
which result in the four equations%
\begin{equation}
\left\{
\begin{array}{c}
\frac{dG}{dR}+\left( \frac{l+2}{R}\right) G+\left( \frac{E}{f_{4}}-M\right)
F=0, \\
\frac{dF}{dR}-\left( \frac{l}{R}\right) F-\left( \frac{E}{f_{4}}+M\right)
G=0,%
\end{array}%
\text{ for }j=l+1/2,\text{ }l=0,1,2,\cdots ,\right.  \label{067}
\end{equation}%
and%
\begin{equation}
\left\{
\begin{array}{c}
\frac{dG}{dR}-\left( \frac{l-1}{R}\right) G+\left( \frac{E}{f_{4}}-M\right)
F=0, \\
\frac{dF}{dR}+\left( \frac{l+1}{R}\right) F-\left( \frac{E}{f_{4}}+M\right)
G=0,%
\end{array}%
\right. \text{ for }j=l-1/2\text{, }l=1,2,3,\cdots \text{.}  \label{068}
\end{equation}%
We leave $1/f_{4}$ in the equations intact since it can be any function of $%
R $ and depends on what kind of interaction we consider. Note that the
constraints, Eqs. (\ref{063})-(\ref{066}), build the connection between a
spacetime structure and a quantum system. By defining the quantum numbers%
\begin{equation}
\kappa =\left\{
\begin{array}{l}
-(j+1/2)=-(l+1)\text{ for }j=l+1/2 \\
+(j+1/2)=l\text{ for }j=l-1/2%
\end{array}%
\right. =\mp 1,\mp 2,\cdots ,  \label{069}
\end{equation}%
two systems (\ref{067}) and (\ref{068}) can be unified and reduced to a
single one for $F$ and $G$%
\begin{equation}
\frac{dF}{dR}+\left( \frac{\kappa +1}{R}\right) F-\left( \frac{E}{f_{4}}%
+M\right) G=0,  \label{070}
\end{equation}%
and%
\begin{equation}
\frac{dG}{dR}-\left( \frac{\kappa -1}{R}\right) G+\left( \frac{E}{f_{4}}%
-M\right) F=0.  \label{071}
\end{equation}%
Together with constraints (\ref{063})-(\ref{066}), the set serves as the
basic equations to obtain a form invariance solution in a central force
field from a spacetime structure.

\section{A spacetime cage for fermions}

As an application of the formulation in section II, we coin a cage of
spacetime for fermions in this section. For this, we choose a trapping zone
with axial symmetry along the $z$ axis. This can be achieved by letting $%
g_{11}=g_{22}$, and thus $f_{1}=f_{2}$. The choice also makes equations (\ref%
{018}), (\ref{019}) and the similar equations in the remaining sets from (%
\ref{023}) to (\ref{062}) are automatically satisfied. As a result, the form
of the spatial components of metric is immediately determined by conditions (%
\ref{064}) and (\ref{065}). They are
\begin{equation}
f_{1}=R^{\prime },\text{ }f_{2}=R^{\prime },\text{ and }f_{3}=\frac{R}{r}.
\label{072}
\end{equation}%
The temporal component of the metric is set to%
\begin{equation}
\frac{1}{f_{4}}=1-\frac{V}{E},  \label{073}
\end{equation}%
with $V$ serving as the potential function, and chosen here to be%
\begin{equation}
V=\left\{
\begin{array}{c}
-V_{0}\text{, \ \ }r\leq r_{b}, \\
0\text{, \ }r>r_{b},%
\end{array}%
\right.  \label{074}
\end{equation}%
where $V_{0}$ is a positive constant. It is obvious that the number $r_{b}$
labels the boundary of the cage. The explicit representation of an
invariance solution needs to know the function $R(r)$. It can be determined
by Eq. (\ref{063}). Since $f_{4}$ is a constant under consideration, it is
not difficult to verify that it is an arbitrary linear function of $r$ under
the conditions (\ref{072}) and (\ref{074}), i.e.,%
\begin{equation}
R(r)=\alpha r+\beta ,  \label{075}
\end{equation}%
where $\alpha $ and $\beta $ are constants. It is time to calculate the form
invariance spinor. By substituting (\ref{073}) into (\ref{070}) and (\ref%
{071}), the form invariant radial equations for region $r\leq r_{b}$ are
given by%
\begin{equation}
\frac{dF}{dR}+\left( \frac{\kappa +1}{R}\right) F-\left( E+V_{0}+M\right)
G=0,  \label{076}
\end{equation}%
and%
\begin{equation}
\frac{dG}{dR}-\left( \frac{\kappa -1}{R}\right) G+\left( E+V_{0}-M\right)
F=0.  \label{077}
\end{equation}%
The equations which $F$ and $R$ satisfy can be found by taking the
derivative of (\ref{076}) and (\ref{077}) with respect to $R$ again, which
gives%
\begin{equation}
\frac{d^{2}F}{dR^{2}}+\frac{2}{R}\frac{dF}{dR}+\left( k^{2}-\frac{\kappa
(\kappa +1)}{R^{2}}\right) F=0,  \label{078}
\end{equation}%
and%
\begin{equation}
\frac{d^{2}G}{dR^{2}}+\frac{2}{R}\frac{dG}{dR}+\left( k^{2}-\frac{\kappa
(\kappa -1)}{R^{2}}\right) G=0,  \label{079}
\end{equation}%
where $k^{2}=(E+V_{0})^{2}-M^{2}$. This is the well-known spherical Bessel
equation. The physical solutions can be divided into two classes. First,
when $k^{2}=(E+V_{0})^{2}-M^{2}>0$, the solution to (\ref{078}) is%
\begin{equation}
F(R)=\left\{
\begin{array}{l}
a_{1}j_{\kappa }(kR)+a_{2}n_{\kappa }(kR),\text{ for }\kappa >0, \\
a_{1}j_{\left\vert \kappa \right\vert -1}(kR)+a_{2}n_{|\kappa |-1}(kR),\text{
for }\kappa <0.%
\end{array}%
\right.  \label{080}
\end{equation}%
Here $a_{1}$ and $a_{2}$ are constants, and $j_{\kappa }$ and $n_{\kappa }$
are the spherical Bessel functions. The solution to (\ref{079}) can be found
through Eq. (\ref{076}) which shows%
\begin{equation}
G=\frac{1}{E+V_{0}+M}\left[ \frac{dF}{dR}+\left( \frac{\kappa +1}{R}\right) F%
\right] .  \label{081}
\end{equation}%
Substituting (\ref{080}) into (\ref{081}) gives rise to the answer%
\begin{equation}
G(R)=\left\{
\begin{array}{l}
\frac{k}{E+V_{0}+M}\left[ a_{1}j_{\kappa -1}(kR)+a_{2}n_{\kappa -1}(kR)%
\right] ,\text{ for }\kappa >0, \\
\frac{-k}{E+V_{0}+M}\left[ a_{1}j_{\left\vert \kappa \right\vert
}(kR)+a_{2}n_{|\kappa |}(kR)\right] ,\text{ for }\kappa <0.%
\end{array}%
\right.  \label{082}
\end{equation}%
To have the representation, we have invoked the recurrence relations%
\begin{equation}
\begin{array}{l}
\frac{n+1}{z}W_{n}(z)+\frac{d}{dz}W_{n}(z)=W_{n-1}(z),\text{ and} \\
\frac{n}{z}W_{n}(z)-\frac{d}{dz}W_{n}(z)=W_{n+1}(z),%
\end{array}
\label{083}
\end{equation}%
where the function $W_{n}(z)$ labels $j_{n}(z)$ and $n_{n}(z)$ (p. 439, \cite%
{6}).

For $k^{2}=(E+V_{0})^{2}-M^{2}<0$, the solution to (\ref{078}) is%
\begin{equation}
F(R)=\left\{
\begin{array}{l}
\sqrt{\frac{2\bar{k}}{\pi R}}\left[ b_{1}I_{\kappa +1/2}(\bar{k}%
R)+b_{2}K_{\kappa +1/2}(\bar{k}R)\right] ,\text{ for }\kappa >0, \\
\sqrt{\frac{2\bar{k}}{\pi R}}\left[ b_{1}I_{\left\vert \kappa \right\vert
-1/2}(\bar{k}R)+b_{2}K_{|\kappa |-1/2}(\bar{k}R)\right] ,\text{ for }\kappa
<0,%
\end{array}%
\right.  \label{084}
\end{equation}%
where we define the symbol $\bar{k}^{2}=M^{2}-(E+V_{0})^{2}>0$, and $I_{\nu
} $ and $K_{\nu }$ are the modified spherical Bessel functions. The
corresponding solution to $G$ is%
\begin{equation}
G(R)=\left\{
\begin{array}{l}
\frac{\bar{k}}{E+V_{0}+M}\sqrt{\frac{2\bar{k}}{\pi R}}\left[ b_{1}I_{\kappa
-1/2}(\bar{k}R)-b_{2}K_{\kappa -1/2}(\bar{k}R)\right] ,\text{ for }\kappa >0,
\\
\frac{\bar{k}}{E+V_{0}+M}\sqrt{\frac{2\bar{k}}{\pi R}}\left[
b_{1}I_{\left\vert \kappa \right\vert +1/2}(\bar{k}R)-b_{2}K_{|\kappa |+1/2}(%
\bar{k}R)\right] ,\text{ for }\kappa <0.%
\end{array}%
\right.  \label{085}
\end{equation}%
The recurrence relations%
\begin{equation}
\begin{array}{l}
\frac{n+1}{z}W_{n}(z)+\frac{d}{dz}W_{n}(z)=W_{n-1}(z),\text{ and} \\
\frac{n}{z}W_{n}(z)-\frac{d}{dz}W_{n}(z)=W_{n+1}(z),%
\end{array}
\label{086}
\end{equation}%
are used to yield the representation in which $W_{n}(z)$ labels $\sqrt{\pi
/2z}I_{n+1/2}(z)$ and $(-1)^{n+1}\sqrt{\pi /2z}K_{n+1/2}(z)$ (p. 444, \cite%
{6}). Using the fact that the wave function has to be continuous, the radial
functions $F(R)$ and $G(R)$ are connected by the functions satisfying the
equations in the region $r>r_{b}$,
\begin{equation}
\frac{dF}{dr}+\left( \frac{\kappa +1}{r}\right) F-\left( E+M\right) G=0,
\label{087}
\end{equation}%
and%
\begin{equation}
\frac{dG}{dr}-\left( \frac{\kappa -1}{r}\right) G+\left( E-M\right) F=0.
\label{088}
\end{equation}%
Obviously, the corresponding solutions to the system are given by%
\begin{equation}
F(r)=\left\{
\begin{array}{l}
a_{1}j_{\kappa }(k_{1}r)+a_{2}n_{\kappa }(k_{1}r),\text{ for }\kappa >0, \\
a_{1}j_{\left\vert \kappa \right\vert -1}(k_{1}r)+a_{2}n_{|\kappa
|-1}(k_{1}r),\text{ for }\kappa <0,%
\end{array}%
\right.  \label{089}
\end{equation}%
and%
\begin{equation}
G(r)=\left\{
\begin{array}{l}
\frac{k_{1}}{E+M}\left[ a_{1}j_{\kappa -1}(k_{1}r)+a_{2}n_{\kappa -1}(k_{1}r)%
\right] ,\text{ for }\kappa >0, \\
\frac{-k_{1}}{E+M}\left[ a_{1}j_{\left\vert \kappa \right\vert
}(k_{1}r)+a_{2}n_{|\kappa |}(k_{1}r)\right] ,\text{ for }\kappa <0,%
\end{array}%
\right.  \label{090}
\end{equation}%
for $k_{1}^{2}=E^{2}-M^{2}>0$, and%
\begin{equation}
F(r)=\left\{
\begin{array}{l}
\sqrt{\frac{2\bar{k}_{1}}{\pi r}}\left[ b_{1}I_{\kappa +1/2}(\bar{k}%
_{1}r)+b_{2}K_{\kappa +1/2}(\bar{k}_{1}r)\right] ,\text{ for }\kappa >0, \\
\sqrt{\frac{2\bar{k}_{1}}{\pi r}}\left[ b_{1}I_{\left\vert \kappa
\right\vert -1/2}(\bar{k}_{1}r)+b_{2}K_{|\kappa |-1/2}(\bar{k}_{1}r)\right] ,%
\text{ for }\kappa <0,%
\end{array}%
\right.  \label{091}
\end{equation}%
and%
\begin{equation}
G(r)=\left\{
\begin{array}{l}
\frac{\bar{k}_{1}}{E+M}\sqrt{\frac{2\bar{k}_{1}}{\pi r}}\left[
b_{1}I_{\kappa -1/2}(\bar{k}_{1}r)-b_{2}K_{\kappa -1/2}(\bar{k}_{1}r)\right]
,\text{ for }\kappa >0, \\
\frac{\bar{k}_{1}}{E+M}\sqrt{\frac{2\bar{k}_{1}}{\pi r}}\left[
b_{1}I_{\left\vert \kappa \right\vert +1/2}(\bar{k}_{1}r)-b_{2}K_{|\kappa
|+1/2}(\bar{k}_{1}r)\right] ,\text{ for }\kappa <0,%
\end{array}%
\right.  \label{092}
\end{equation}%
for $\bar{k}_{1}=M^{2}-E^{2}>0$.

\subsection{Bound states}

Let us now discuss the bound states caught by the spacetime cage. For
fermions with energy satisfying the condition $k^{2}=(E+V_{0})^{2}-M^{2}>0$,
their radial amplitudes in the cage are described by (\ref{080}) and (\ref%
{082}) with the coefficient $a_{2}=0$ since $n_{n}(z)\longrightarrow
1/z^{n+1}$ as $z\rightarrow 0$. Outside the cage, the representations (\ref%
{091}) and (\ref{092}) with $b_{1}=0$ are the radial solutions of fermions
since the wave functions are bounded, and $I_{n+1/2}(z)\rightarrow e^{z}/%
\sqrt{z}$ as $z\rightarrow \infty $. The energy levels of the states are
determined by the continuous condition of wave functions at the boundary $%
r=r_{b}$. It is straightforward to show that the spectrum is determined by
the equation%
\begin{equation}
\frac{j_{l_{1}}(kr_{b})}{j_{l_{2}}(kr_{b})}=-\epsilon _{\kappa }\frac{k}{%
\bar{k}_{1}}\left( \frac{E+M}{E+V_{0}+M}\right) \frac{K_{l_{1}+1/2}(\bar{k}%
_{1}r_{b})}{K_{l_{2}+1/2}(\bar{k}_{1}r_{b})},  \label{093}
\end{equation}%
where $\epsilon _{\kappa }=1$ $(-1)$ for $\kappa >0$ $(\kappa <0)$, and the
symbols $l_{1}$ and $l_{2}$ denote%
\begin{equation}
l_{1}=\left\{
\begin{array}{l}
\kappa \text{, for }\kappa >0, \\
|\kappa |-1\text{, for }\kappa <0,%
\end{array}%
\right. \text{ \ and }l_{2}=\left\{
\begin{array}{l}
\kappa -1\text{, for }\kappa >0, \\
|\kappa |\text{, for }\kappa <0.%
\end{array}%
\right.  \label{094}
\end{equation}%
For $s$ states, the quantum number $\kappa =-1$. The energy levels satisfy
the relation%
\begin{equation}
\frac{j_{0}(kr_{b})}{j_{1}(kr_{b})}=\frac{k}{\bar{k}_{1}}\left( \frac{E+M}{%
E+V_{0}+M}\right) \frac{K_{1/2}(\bar{k}_{1}r_{b})}{K_{3/2}(\bar{k}_{1}r_{b})}%
.  \label{095}
\end{equation}%
Fig. 1 shows the energy spectrum allowed by the $s$ states. The boundaries
of the cage are taken as $r_{b}=10\lambda _{F}$ and $100\lambda _{F}$ with $%
\lambda _{F}=\hbar /Mc$, the Compton wavelength of fermions. The spectrum
equation (\ref{093}) is actually the same as that in the usual spherical box
for fermions \cite{6a}. However, the presented spacetime approach to the
quantization rule has more implications hidden in the freedom of solutions
with the form invariance. In equation (\ref{075}), we have shown that the
radial amplitude of fermions in the cage is with the function $R(r)=\alpha
r+\beta $ as a variable. One can choose%
\begin{equation}
\alpha =\frac{r_{b}}{r_{b}-r_{a}}\text{, and }\beta =\frac{-r_{a}r_{b}}{%
r_{b}-r_{a}}.  \label{096}
\end{equation}%
This choice implies that the spacetime cage can capture fermion states
accompanying an invisible region. Fig. 2 shows the component $F(R(r))$ for
the $1s_{1/2}$ state in the cage. It demonstrates the zone sustaining a
fermion state accompanying a zero amplitude invisible region.

\subsection{Scattering states}

The presented cage can be regarded as a quantum cloak for bound states since
it establishes an invisible region for them and does not alter the behavior
of scattering states. For completeness, the scattering behavior of fermions
from the cage is discussed in this subsection. The allowed physical solution
in the cage is given by%
\begin{eqnarray}
F^{i}(R) &=&a_{1}j_{l_{1}}(kR),  \label{097} \\
G^{i}(R) &=&\epsilon _{\kappa }\frac{k}{E+V_{0}+M}a_{1}j_{l_{2}}(kR),
\nonumber
\end{eqnarray}%
for every $\kappa $. Outside the cage, the scattering radial functions are%
\begin{eqnarray}
F^{o}(r) &=&c_{1}j_{l_{1}}(k_{1}r)+c_{2}n_{l_{1}}(k_{1}r),  \label{098} \\
G^{o}(r) &=&\epsilon _{\kappa }\frac{k_{1}}{E+M}\left[
c_{1}j_{l_{2}}(k_{1}r)+c_{2}n_{l_{2}}(k_{1}r)\right] ,  \nonumber
\end{eqnarray}%
where $k_{1}^{2}=E^{2}-M^{2}$. An asymptotic analysis of the outgoing wave
can be made with the formulas $j_{\nu }(z)\rightarrow \sin (z-\nu \pi /2)/z$
and $n_{\nu }(z)\rightarrow -\cos (z-\nu \pi /2)/z$ as $z\rightarrow \infty $%
. Consequently, the radial parts of the outgoing wave in the far zone are
given by
\begin{eqnarray}
F^{o}(r) &\longrightarrow &\frac{A}{k_{1}r}\sin (k_{1}r-l_{1}\pi /2+\delta
_{\kappa }),  \label{099} \\
G^{o}(r) &\longrightarrow &\epsilon _{\kappa }\frac{k_{1}}{E+M}\frac{A}{%
k_{1}r}\sin (k_{1}r-l_{2}\pi /2+\delta _{\kappa }),  \nonumber
\end{eqnarray}%
where the phase shift is defined by $c_{2}/c_{1}=-\tan \delta _{\kappa }$,
and $A=c_{1}/\cos \delta _{\kappa }$. Using the fact that the spinor is
continuous at $r=r_{b}$,%
\begin{equation}
c_{1}j_{l_{1}}(k_{1}r_{b})+c_{2}n_{l_{1}}(k_{1}r_{b})=a_{1}j_{l_{1}}(kr_{b}),
\label{0100}
\end{equation}%
and

\begin{equation}
\epsilon _{\kappa }\frac{k_{1}}{E+M}\left[
c_{1}j_{l_{2}}(k_{1}r_{b})+c_{2}n_{l_{2}}(k_{1}r_{b})\right] =\epsilon
_{\kappa }\frac{k}{E+V_{0}+M}a_{1}j_{l_{2}}(kr_{b}).  \label{0101}
\end{equation}%
It is easy to show that the phase shift has the analytical representation%
\begin{equation}
\tan \delta _{\kappa }=\frac{%
j_{l_{1}}(kr_{b})j_{l_{2}}(k_{1}r_{b})-Bj_{l_{1}}(k_{1}r_{b})j_{l_{2}}(kr_{b})%
}{%
j_{l_{1}}(kr_{b})n_{l_{2}}(k_{1}r_{b})-Bj_{l_{2}}(kr_{b})n_{l_{1}}(k_{1}r_{b})%
},  \label{0102}
\end{equation}%
where the constant
\begin{equation}
B=\sqrt{\frac{\left( E+M\right) \left( E+V_{0}-M\right) }{\left( E-M\right)
\left( E+V_{0}+M\right) }}.  \label{0103}
\end{equation}%
A phase shift often reflects the subtle structure of the bound states \cite{lin2, lin3}.
Fig. 3 shows the phase shift of the $s$ states for $V_{0}=2Mc^{2}$ and radius $%
r_{b}=10\lambda _{F}$, where the phase shift has absorbed $k_{1}r_{b}$,
i.e. $\delta=\delta _{\kappa=-1}+k_{1}r_{b}$,
since it can show that all $s$ states have the factor.

\subsection{Experiment}

We here explain how it is possible to coin the trapping cage with a physical
method. Eqs. (\ref{072}) and (\ref{073}) show that the cage is created by a
curved space depicted by the line element%
\begin{equation}
ds^{2}=\alpha ^{2}(dx^{1})^{2}+\alpha ^{2}(dx^{2})^{2}+\left( \alpha +\frac{%
\beta }{r}\right) ^{2}(dx^{3})^{2}-\left( \frac{1}{1+V_{0}/E}\right)
^{2}(dt)^{2}.  \label{0104}
\end{equation}%
Constant $\alpha $ and $\beta $ are as in (\ref{096}). Note that there is
the same proper time on every point in the cage. This is easy to be seen by%
\begin{equation}
d\tau =ids=\sqrt{g_{44}}dt=\left( \frac{1}{1+V_{0}/E}\right) dt.
\label{0104a}
\end{equation}%
The element of spatial distance for all space is given by (p. 235, \cite{7})%
\begin{equation}
dL^{2}=\left( g_{ij}-\frac{g_{0i}g_{0j}}{g_{00}}\right) dx^{i}dx^{j}\text{,
\ }i,j=1,2,3.  \label{0105}
\end{equation}%
Since there are no off diagonal components here, the element of spatial
distance has the simple representation%
\begin{equation}
dL^{2}=\alpha ^{2}(dx^{1})^{2}+\alpha ^{2}(dx^{2})^{2}+\left( \alpha +\frac{%
\beta }{r}\right) ^{2}(dx^{3})^{2}.  \label{0106}
\end{equation}%
The first step to the cage is to coin a spatial zone with the line element $%
dL^{2}$. In a laboratory, the interval of both space and time for the zone
is given by%
\begin{equation}
ds^{2}=\alpha ^{2}(dx^{1})^{2}+\alpha ^{2}(dx^{2})^{2}+\left( \alpha +\frac{%
\beta }{r}\right) ^{2}(dx^{3})^{2}-(dt)^{2}.  \label{0107}
\end{equation}%
Here $t$ is defined by a running cloak in the laboratory. It shows that the
elapsed time on each point of the zone is the same as the laboratory. To
have the effect of the proper time shown as (\ref{0104a}), one pushes the 3D
trapping zone along the $x^{1}$ or $x^{2}$ axis, say $x^{1}$, to have a
constant velocity $v$. Eq. (\ref{0107}) shows that%
\begin{equation}
ds^{2}=\left[ \alpha ^{2}\left( \frac{dx^{1}}{dt}\right) ^{2}-1\right]
(dt)^{2}=\left( \alpha ^{2}v^{2}-1\right) (dt)^{2},  \label{0123}
\end{equation}%
that is,%
\begin{equation}
d\tau =ids=\sqrt{1-\alpha ^{2}v^{2}}dt.  \label{0124}
\end{equation}%
A comparison of this equation with (\ref{0104a}) gives%
\begin{equation}
\sqrt{1-\alpha ^{2}v^{2}}=\left( \frac{E}{E+V_{0}}\right) .  \label{0125}
\end{equation}%
Therefore, the velocity to create the constant global proper time in the
cage is%
\begin{equation}
v=\frac{c}{\alpha }\sqrt{1-\left( \frac{E}{E+V_{0}}\right) ^{2}},
\label{0126}
\end{equation}%
where we put the light speed $c$ back clearly. According to the numerical
result of Fig. 1, for the cage with radius $r_{b}=100\lambda _{F}$, the
first bound state $1s_{1/2}$ appears at $V_{0}/E_{0}=0.008059$ with energy $%
E/E_{0}=0.992$. The corresponding velocity in (\ref{0126}) is about $%
v=0.1267c/\alpha $. A large number $\alpha $ will greatly reduce the
velocity. The constant velocity of the zone can actually also be offered by
a circular motion. The Lorentz contraction due to the velocity can be
ignored here since it is less than 1\% for the maximum $v=0.1267c$. Before
closing the section, let's make two notes on the asymptotic behavior of the
cage. First, the parameter $\alpha =1$ when $r_{a}=0$. The spacetime
structure becomes%
\begin{equation}
ds^{2}=(dx^{1})^{2}+(dx^{2})^{2}+(dx^{3})^{2}-\left( \frac{1}{1+V_{0}/E}%
\right) ^{2}(dt)^{2}.  \label{0127}
\end{equation}%
Therefore, a pure curved structure of time with $g_{44}=\left[ E/(E+V_{0})%
\right] ^{2}$suffices to obtain the trapping effect of an ideal potential
well for fermions. However, as stated above, the lowest velocity to have the
effect is about 13\% of the light speed. Second, the parameter $\alpha
\rightarrow \infty $ when $r_{a}\rightarrow r_{b}$. The velocity $v$ for the
cage is easier to reach. However, the spatial geometry becomes complicated.
A balance between the velocity which we can push and the difficulty of
making a space with the metric (\ref{0106}) should be estimated.

\section{CONCLUSION}

Constructing the connection between a spacetime structure and a quantum rule
has long attracted considerable attention since the general invariance
principle was established \cite{8}. In this paper, we give the constraint
conditions for the covariant Dirac equation which determines the spacetime
structure of constructing a quantum system with the form invariance solution
in the central force field. As an application, constructing a spacetime cage
that confines fermions forming bound states and builds the quantum rules of\
a spherical well is presented. Let us remark on several consequences of our
discussion: (i) The quantum rule of fermions created by a spacetime
structure accompanies novel fermion states. As shown in Sec. III, the
quantum rule of fermions in a spherical well can be built from a spacetime
geometry, while the corresponding states accompany an invisible region.
Obviously, any quantum rules using the spacetime approach correspond to the
new type of fermion states due to the form invariance variable. (ii) There
may be a chance to design a new quantum device with a macroscopic scale
using the presented spacetime approach. In general, the curved structure of
a space is often involved in the discussion of the motion of macroscopic
objects. Establishing the connection between a quantum system and a metric
structure makes the construction of a macroscopic quantum system possible.
For instance, a neutrino mass is less than 2eV \cite{8a}. The corresponding
Compton wavelength $\lambda _{{\rm neutrino}}>10^{-5}cm$. Making a spacetime
cage for neutrinos with radius $r_{b}=500\lambda _{{\rm neutrino}}$ could
have the width $w=10^{3}\lambda _{{\rm neutrino}}>10^{-2}cm$. (iii) The
trapping cage coined from the spacetime structure may serve as a bag to get
the bizarre spin-1/2 particles. The trapping force emerging from our
discussion is constructed from the shape of spacetime. It is applicable to
any spin-1/2 particles. The approach here may be a ladder to get to the
trapping job for spin-1/2 dark particles. Specifically, relic neutrinos have
very low energy, which carry information of our universe about one second
old. The presented construct may be applicable to catch these marvelous
particles.

\newpage \appendix

\section{\newline
Proof of Eqs. (\protect\ref{018})-(\protect\ref{021}) and (\protect\ref{023}%
)-(\protect\ref{026})}

\label{App:AppendixA}

In deriving the form invariance Dirac spinors, we need the formulas for the
action of the coordinates on the spherical harmonics. We list them as
follows:
\begin{equation}
\frac{2x}{r}Y_{lm}=\sqrt{\frac{(l-m-1)(l-m)}{(2l+1)(2l-1)}}Y_{l-1,m+1}-\sqrt{%
\frac{(l+m+2)(l+m+1)}{(2l+3)(2l+1)}}Y_{l+1,m+1}  \label{A1}
\end{equation}%
\[
-\sqrt{\frac{(l+m)(l+m-1)}{(2l+1)(2l-1)}}Y_{l-1,m-1}+\sqrt{\frac{%
(l-m+2)(l-m+1)}{(2l+3)(2l+1)}}Y_{l+1,m-1},
\]%
\begin{equation}
\frac{2iy}{r}Y_{lm}=\sqrt{\frac{(l-m-1)(l-m)}{(2l+1)(2l-1)}}Y_{l-1,m+1}-%
\sqrt{\frac{(l+m+2)(l+m+1)}{(2l+3)(2l+1)}}Y_{l+1,m+1}  \label{A2}
\end{equation}%
\[
+\sqrt{\frac{(l+m)(l+m-1)}{(2l+1)(2l-1)}}Y_{l-1,m-1}-\sqrt{\frac{%
(l-m+2)(l-m+1)}{(2l+3)(2l+1)}}Y_{l+1,m-1},
\]%
and%
\begin{equation}
\frac{z}{r}Y_{lm}=\sqrt{\frac{(l+m+1)(l-m+1)}{(2l+3)(2l+1)}}Y_{l+1,m}+\sqrt{%
\frac{(l+m)(l-m)}{(2l+1)(2l-1)}}Y_{l-1,m}.  \label{A3}
\end{equation}%
Also, derivatives of the wave function with respect to the coordinates
occur. They have relations (see, e.g., pp. 346-349 \cite{9})%
\[
\left( \frac{\partial }{\partial x}+i\frac{\partial }{\partial y}\right)
\left( GY_{lm}\right) =\sqrt{\frac{(l+m+2)(l+m+1)}{(2l+3)(2l+1)}}%
Y_{l+1,m+1}\left( \frac{dG}{dr}-l\frac{G}{r}\right)
\]%
\begin{equation}
-\sqrt{\frac{(l-m)(l-m-1)}{(2l+1)(2l-1)}}Y_{l-1,m+1}\left( \frac{dG}{dr}%
+(l+1)\frac{G}{r}\right) ,  \label{A4}
\end{equation}%
\[
\left( \frac{\partial }{\partial x}-i\frac{\partial }{\partial y}\right)
\left( GY_{lm}\right) =-\sqrt{\frac{(l-m+2)(l-m+1)}{(2l+3)(2l+1)}}%
Y_{l+1,m-1}\left( \frac{dG}{dr}-l\frac{G}{r}\right)
\]%
\begin{equation}
+\sqrt{\frac{(l+m)(l+m-1)}{(2l+1)(2l-1)}}Y_{l-1,m-1}\left( \frac{dG}{dr}%
+(l+1)\frac{G}{r}\right) ,  \label{A5}
\end{equation}%
and%
\[
\frac{\partial }{\partial z}\left( GY_{lm}\right) =\sqrt{\frac{(l+m+1)(l-m+1)%
}{(2l+3)(2l+1)}}Y_{l+1,m}\left( \frac{dG}{dr}-l\frac{G}{r}\right)
\]%
\begin{equation}
+\sqrt{\frac{(l+m)(l-m)}{(2l+1)(2l-1)}}Y_{l-1,m}\left( \frac{dG}{dr}+(l+1)%
\frac{G}{r}\right) .  \label{A6}
\end{equation}%
With the relations (\ref{A1})-(\ref{A3}), we get%
\[
\frac{2x}{r}Y_{l+1,m+1/2}=\sqrt{\frac{(l-m-1/2)(l-m+1/2)}{(2l+3)(2l+1)}}%
Y_{l,m+3/2}-\sqrt{\frac{(l+m+7/2)(l+m+5/2)}{(2l+5)(2l+3)}}Y_{l+2,m+3/2}
\]%
\begin{equation}
-\sqrt{\frac{(l+m+3/2)(l+m+1/2)}{(2l+3)(2l+1)}}Y_{l,m-1/2}+\sqrt{\frac{%
(l-m+5/2)(l-m+3/2)}{(2l+5)(2l+3)}}Y_{l+2,m-1/2},  \label{A7}
\end{equation}%
\[
\frac{2iy}{r}Y_{l+1,m+1/2}=\sqrt{\frac{(l-m-1/2)(l-m+1/2)}{(2l+3)(2l+1)}}%
Y_{l,m+3/2}-\sqrt{\frac{(l+m+7/2)(l+m+5/2)}{(2l+5)(2l+3)}}Y_{l+2,m+3/2}
\]%
\begin{equation}
+\sqrt{\frac{(l+m+3/2)(l+m+1/2)}{(2l+3)(2l+1)}}Y_{l,m-1/2}-\sqrt{\frac{%
(l-m+5/2)(l-m+3/2)}{(2l+5)(2l+3)}}Y_{l+2,m-1/2},  \label{A8}
\end{equation}%
and%
\begin{equation}
\frac{z}{r}Y_{l+1,m-1/2}=\sqrt{\frac{(l+m+3/2)(l-m+5/2)}{(2l+5)(2l+3)}}%
Y_{l+2,m-1/2}+\sqrt{\frac{(l+m+1/2)(l-m+3/2)}{(2l+3)(2l+1)}}Y_{l,m-1/2}.
\label{A9}
\end{equation}%
With the relations (\ref{A4})-(\ref{A6}), we have%
\[
2\frac{\partial }{\partial x}\left( GY_{l+1,m+1/2}\right) =\sqrt{\frac{%
(l+m+7/2)(l+m+5/2)}{(2l+5)(2l+3)}}Y_{l+2,m+3/2}\left( \frac{dG}{dr}-\left(
\frac{l+1}{r}\right) G\right)
\]%
\[
-\sqrt{\frac{(l-m+1/2)(l-m-1/2)}{(2l+3)(2l+1)}}Y_{l,m+3/2}\left( \frac{dG}{dr%
}+\left( \frac{l+2}{r}\right) G\right)
\]%
\[
-\sqrt{\frac{(l-m+5/2)(l-m+3/2)}{(2l+5)(2l+3)}}Y_{l+2,m-1/2}\left( \frac{dG}{%
dr}-\left( \frac{l+1}{r}\right) G\right)
\]%
\begin{equation}
+\sqrt{\frac{(l+m+3/2)(l+m+1/2)}{(2l+3)(2l+1)}}Y_{l,m-1/2}\left( \frac{dG}{dr%
}+\left( \frac{l+2}{r}\right) G\right) ,  \label{A10}
\end{equation}%
\[
2i\frac{\partial }{\partial y}\left( GY_{l+1,m+1/2}\right) =\sqrt{\frac{%
(l+m+7/2)(l+m+5/2)}{(2l+5)(2l+3)}}Y_{l+2,m+3/2}\left( \frac{dG}{dr}-\left(
\frac{l+1}{r}\right) G\right)
\]%
\[
-\sqrt{\frac{(l-m+1/2)(l-m-1/2)}{(2l+3)(2l+1)}}Y_{l,m+3/2}\left( \frac{dG}{dr%
}+\left( \frac{l+2}{r}\right) G\right)
\]%
\[
+\sqrt{\frac{(l-m+5/2)(l-m+3/2)}{(2l+5)(2l+3)}}Y_{l+2,m-1/2}\left( \frac{dG}{%
dr}-\left( \frac{l+1}{r}\right) G\right)
\]%
\begin{equation}
-\sqrt{\frac{(l+m+3/2)(l+m+1/2)}{(2l+3)(2l+1)}}Y_{l,m-1/2}\left( \frac{dG}{dr%
}+\left( \frac{l+2}{r}\right) G\right) ,  \label{A11}
\end{equation}%
and%
\[
\frac{\partial }{\partial z}\left( GY_{l+1,m-1/2}\right) =\sqrt{\frac{%
(l+m+3/2)(l-m+5/2)}{(2l+5)(2l+3)}}Y_{l+2,m-1/2}\left( \frac{dG}{dr}-\left(
\frac{l+1}{r}\right) G\right)
\]%
\begin{equation}
+\sqrt{\frac{(l+m+1/2)(l-m+3/2)}{(2l+3)(2l+1)}}Y_{l,m-1/2}\left( \frac{dG}{dr%
}+\left( \frac{l+2}{r}\right) G\right) .  \label{A12}
\end{equation}%
Apply (\ref{A7})-(\ref{A12}) to (\ref{012}) for the trial solution (\ref{016}%
). Eq. (\ref{012}) has the representation separated by the spherical
harmonics%
\[
-\frac{1}{2(2l+3)}\sqrt{\frac{(l+m+7/2)(l+m+5/2)(l+m+3/2)}{(2l+5)}}\left(
\frac{1}{f_{1}}\right) Y_{l+2,m+3/2}\left( \frac{dG}{dr}-\left( \frac{l+1}{r}%
\right) G\right)
\]%
\[
+\frac{1}{2(2l+3)}\sqrt{\frac{(l+m+3/2)(l-m+1/2)(l-m-1/2)}{(2l+1)}}\left(
\frac{1}{f_{1}}\right) Y_{l,m+3/2}\left( \frac{dG}{dr}+\left( \frac{l+2}{r}%
\right) G\right)
\]%
\[
+\frac{1}{2(2l+3)}\sqrt{\frac{(l+m+3/2)(l-m+5/2)(l-m+3/2)}{(2l+5)}}\left(
\frac{1}{f_{1}}\right) Y_{l+2,m-1/2}\left( \frac{dG}{dr}-\left( \frac{l+1}{r}%
\right) G\right)
\]%
\[
-\frac{1}{2(2l+3)}\sqrt{\frac{(l+m+3/2)^{2}(l+m+1/2)}{(2l+1)}}\left( \frac{1%
}{f_{1}}\right) Y_{l,m-1/2}\left( \frac{dG}{dr}+\left( \frac{l+2}{r}\right)
G\right)
\]%
\[
+\frac{1}{2(2l+3)}\sqrt{\frac{(l+m+3/2)(l-m+1/2)(l-m-1/2)}{(2l+1)}}\left(
\frac{R^{\prime }}{2f_{1}}\right) GY_{l,m+3/2}\Delta _{1}
\]%
\[
-\frac{1}{2(2l+3)}\sqrt{\frac{(l+m+7/2)(l+m+5/2)(l+m+3/2)}{(2l+5)}}\left(
\frac{R^{\prime }}{2f_{1}}\right) GY_{l+2,m+3/2}\Delta _{1}
\]%
\[
-\frac{1}{2(2l+3)}\sqrt{\frac{(l+m+3/2)^{2}(l+m+1/2)}{(2l+1)}}\left( \frac{%
R^{\prime }}{2f_{1}}\right) GY_{l,m-1/2}\Delta _{1}
\]%
\[
+\frac{1}{2(2l+3)}\sqrt{\frac{(l+m+3/2)(l-m+5/2)(l-m+3/2)}{(2l+5)}}\left(
\frac{R^{\prime }}{2f_{1}}\right) GY_{l+2,m-1/2}\Delta _{1}
\]%
\[
+\frac{1}{2(2l+3)}\sqrt{\frac{(l+m+7/2)(l+m+5/2)(l+m+3/2)}{(2l+5)}}\left(
\frac{1}{f_{2}}\right) Y_{l+2,m+3/2}\left( \frac{dG}{dr}-\left( \frac{l+1}{r}%
\right) G\right)
\]%
\[
-\frac{1}{2(2l+3)}\sqrt{\frac{(l+m+3/2)(l-m+1/2)(l-m-1/2)}{(2l+1)}}\left(
\frac{1}{f_{2}}\right) Y_{l,m+3/2}\left( \frac{dG}{dr}+\left( \frac{l+2}{r}%
\right) G\right)
\]%
\[
+\frac{1}{2(2l+3)}\sqrt{\frac{(l+m+3/2)(l-m+5/2)(l-m+3/2)}{(2l+5)}}\left(
\frac{1}{f_{2}}\right) Y_{l+2,m-1/2}\left( \frac{dG}{dr}-\left( \frac{l+1}{r}%
\right) G\right)
\]%
\[
-\frac{1}{2(2l+3)}\sqrt{\frac{(l+m+3/2)^{2}(l+m+1/2)}{(2l+1)}}\left( \frac{1%
}{f_{2}}\right) Y_{l,m-1/2}\left( \frac{dG}{dr}+\left( \frac{l+2}{r}\right)
G\right)
\]%
\[
-\frac{1}{2(2l+3)}\sqrt{\frac{(l+m+3/2)(l-m+1/2)(l-m-1/2)}{(2l+1)}}\left(
\frac{R^{\prime }}{2f_{2}}\right) GY_{l,m+3/2}\Delta _{2}
\]%
\[
+\frac{1}{2(2l+3)}\sqrt{\frac{(l+m+7/2)(l+m+5/2)(l+m+3/2)}{(2l+5)}}\left(
\frac{R^{\prime }}{2f_{2}}\right) GY_{l+2,m+3/2}\Delta _{2}
\]%
\[
-\frac{1}{2(2l+3)}\sqrt{\frac{(l+m+3/2)^{2}(l+m+1/2)}{(2l+1)}}\left( \frac{%
R^{\prime }}{2f_{2}}\right) GY_{l,m-1/2}\Delta _{2}
\]%
\[
+\frac{1}{2(2l+3)}\sqrt{\frac{(l+m+3/2)(l-m+5/2)(l-m+3/2)}{(2l+5)}}\left(
\frac{R^{\prime }}{2f_{2}}\right) GY_{l+2,m-1/2}\Delta _{2}
\]%
\[
-\frac{1}{(2l+3)}\sqrt{\frac{(l+m+3/2)(l-m+5/2)(l-m+3/2)}{(2l+5)}}\left(
\frac{1}{f_{3}}\right) Y_{l+2,m-1/2}\left( \frac{dG}{dr}-\left( \frac{l+1}{r}%
\right) G\right)
\]%
\[
-\frac{1}{(2l+3)}\sqrt{\frac{(l+m+1/2)(l-m+3/2)^{2}}{(2l+1)}}\left( \frac{1}{%
f_{3}}\right) Y_{l,m-3/2}\left( \frac{dG}{dr}+\left( \frac{l+2}{r}\right)
G\right)
\]%
\[
+\frac{1}{(2l+3)}\sqrt{\frac{(l+m+3/2)(l-m+5/2)(l-m+3/2)}{(2l+5)}}\left(
\frac{R^{\prime }}{2f_{3}}\right) GY_{l+2,m-3/2}\Delta _{3}
\]%
\[
+\frac{1}{(2l+3)}\sqrt{\frac{(l+m+1/2)(l-m+3/2)^{2}}{(2l+1)}}\left( \frac{%
R^{\prime }}{2f_{3}}\right) GY_{l,m-1/2}\Delta _{3}
\]%
\begin{equation}
-\left( \frac{E}{f_{4}}-M\right) \sqrt{\frac{(l+m+1/2)}{(2l+1)}}%
FY_{l,m-1/2}=0,  \label{A13}
\end{equation}%
where $\Delta _{i}$ are as defined in (\ref{022}). Multiplying this equation
by $Y_{l+2,m+3/2}$, and integrating both sides of it with respect to the
solid angle $d\Omega =\sin \theta d\theta d\varphi $ gives the equality%
\begin{equation}
\left( -\frac{1}{f_{1}}+\frac{1}{f_{2}}\right) \left[ \frac{dG}{dr}-\left(
\frac{l+1}{r}\right) G\right] +\left[ -\frac{(\Delta _{1})}{f_{1}}+\frac{%
(\Delta _{2})}{f_{2}}\right] \frac{R^{\prime }}{2}G=0.  \label{A14}
\end{equation}%
Multiplying both sides by $Y_{l,m+3/2}$, and performing integration with
respect to the solid angle gets

\begin{equation}
\left( \frac{1}{f_{1}}-\frac{1}{f_{2}}\right) \left[ \frac{dG}{dr}+\left(
\frac{l+2}{r}\right) G\right] +\left[ \frac{(\Delta _{1})}{f_{1}}-\frac{%
(\Delta _{2})}{f_{2}}\right] \frac{R^{\prime }}{2}G=0.  \label{A15}
\end{equation}%
Multiplying both sides by $Y_{l+2,m-1/2}$, and performing the integration as
above, it follows that%
\begin{equation}
\left( \frac{1}{2f_{1}}+\frac{1}{2f_{2}}-\frac{1}{f_{3}}\right) \left[ \frac{%
dG}{dr}-\left( \frac{l+1}{r}\right) G\right] +\left[ \frac{(\Delta _{1})}{%
2f_{1}}+\frac{(\Delta _{2})}{2f_{2}}+\frac{(\Delta _{3})}{f_{3}}\right]
\frac{R^{\prime }}{2}G=0.  \label{A16}
\end{equation}%
Multiplying both sides by $Y_{l,m-1/2}$, and performing the integration of
the solid angle results in%
\[
-\frac{l+m+3/2}{2(2l+3)}\left\{ \left( \frac{1}{f_{1}}+\frac{1}{f_{2}}%
\right) \left[ \frac{dG}{dr}+\left( \frac{l+2}{r}\right) G\right] +\left[
\frac{(\Delta _{1})}{f_{1}}+\frac{(\Delta _{2})}{f_{2}}\right] \frac{%
R^{\prime }}{2}G\right\}
\]%
\begin{equation}
+\left( \frac{l-m+3/2}{2l+3}\right) \left\{ -\frac{1}{f_{3}}\left[ \frac{dG}{%
dr}+\left( \frac{l+2}{r}\right) G\right] +\frac{(\Delta _{3})}{f_{3}}\frac{%
R^{\prime }}{2}G\right\} +\left( -\frac{E}{f_{4}}+M\right) F=0.  \label{A17}
\end{equation}%
Eqs. (\ref{A14})-(\ref{A17}) are Eqs. (\ref{018})-(\ref{021}). The proof is
completed.

We turn to the proof of the equivalence between (\ref{013}) and (\ref{023})-(%
\ref{026}). It follows from the recurrence relations in (\ref{A1})-(\ref{A3}%
) that
\[
\frac{2x}{r}Y_{l+1,m-1/2}=\sqrt{\frac{(l-m+3/2)(l-m+1/2)}{(2l+3)(2l+1)}}%
Y_{l,m+1/2}-\sqrt{\frac{(l+m+5/2)(l+m+3/2)}{(2l+5)(2l+3)}}Y_{l+2,m+1/2}
\]%
\begin{equation}
-\sqrt{\frac{(l+m+1/2)(l+m-1/2)}{(2l+3)(2l+1)}}Y_{l,m-3/2}+\sqrt{\frac{%
(l-m+7/2)(l-m+5/2)}{(2l+5)(2l+3)}}Y_{l+2,m-3/2},  \label{A18}
\end{equation}%
\[
\frac{2iy}{r}Y_{l+1,m-1/2}=\sqrt{\frac{(l-m+3/2)(l-m+1/2)}{(2l+3)(2l+1)}}%
Y_{l,m+1/2}-\sqrt{\frac{(l+m+5/2)(l+m+3/2)}{(2l+5)(2l+3)}}Y_{l+2,m+1/2}
\]%
\begin{equation}
+\sqrt{\frac{(l+m+1/2)(l+m-1/2)}{(2l+3)(2l+1)}}Y_{l,m-3/2}-\sqrt{\frac{%
(l-m+7/2)(l-m+5/2)}{(2l+5)(2l+3)}}Y_{l+2,m-3/2},  \label{A19}
\end{equation}%
and%
\begin{equation}
\frac{z}{r}Y_{l+1,m+1/2}=\sqrt{\frac{(l+m+5/2)(l-m+3/2)}{(2l+5)(2l+3)}}%
Y_{l+2,m+1/2}+\sqrt{\frac{(l+m+3/2)(l-m+1/2)}{(2l+3)(2l+1)}}Y_{l,m+1/2}.
\label{A20}
\end{equation}%
With the relations (\ref{A4})-(\ref{A6}), we have%
\[
2\frac{\partial }{\partial x}\left( GY_{l+1,m-1/2}\right) =\sqrt{\frac{%
(l+m+5/2)(l+m+3/2)}{(2l+5)(2l+3)}}Y_{l+2,m+1/2}\left( \frac{dG}{dr}-\left(
\frac{l+1}{r}\right) G\right)
\]%
\[
-\sqrt{\frac{(l-m+3/2)(l-m+1/2)}{(2l+3)(2l+1)}}Y_{l,m+1/2}\left( \frac{dG}{dr%
}+\left( \frac{l+2}{r}\right) G\right)
\]%
\[
-\sqrt{\frac{(l-m+7/2)(l-m+5/2)}{(2l+5)(2l+3)}}Y_{l+2,m-3/2}\left( \frac{dG}{%
dr}-\left( \frac{l+1}{r}\right) G\right)
\]%
\begin{equation}
+\sqrt{\frac{(l+m+1/2)(l+m-1/2)}{(2l+3)(2l+1)}}Y_{l,m-3/2}\left( \frac{dG}{dr%
}+\left( \frac{l+2}{r}\right) G\right) ,  \label{A21}
\end{equation}%
\[
2i\frac{\partial }{\partial y}\left( GY_{l+1,m-1/2}\right) =\sqrt{\frac{%
(l+m+5/2)(l+m+3/2)}{(2l+5)(2l+3)}}Y_{l+2,m+1/2}\left( \frac{dG}{dr}-\left(
\frac{l+1}{r}\right) G\right)
\]%
\[
-\sqrt{\frac{(l-m+3/2)(l-m+1/2)}{(2l+3)(2l+1)}}Y_{l,m+1/2}\left( \frac{dG}{dr%
}+\left( \frac{l+2}{r}\right) G\right)
\]%
\[
+\sqrt{\frac{(l-m+7/2)(l-m+5/2)}{(2l+5)(2l+3)}}Y_{l+2,m-3/2}\left( \frac{dG}{%
dr}-\left( \frac{l+1}{r}\right) G\right)
\]%
\begin{equation}
-\sqrt{\frac{(l+m+1/2)(l+m-1/2)}{(2l+3)(2l+1)}}Y_{l,m-3/2}\left( \frac{dG}{dr%
}+\left( \frac{l+2}{r}\right) G\right) ,  \label{A22}
\end{equation}%
and%
\[
\frac{\partial }{\partial z}\left( GY_{l+1,m+1/2}\right) =\sqrt{\frac{%
(l+m+5/2)(l-m+3/2)}{(2l+5)(2l+3)}}Y_{l+2,m+1/2}\left( \frac{dG}{dr}-\left(
\frac{l+1}{r}\right) G\right)
\]%
\begin{equation}
+\sqrt{\frac{(l+m+3/2)(l-m+1/2)}{(2l+3)(2l+1)}}Y_{l,m+1/2}\left( \frac{dG}{dr%
}+\left( \frac{l+2}{r}\right) G\right) .  \label{A23}
\end{equation}%
Substitution of (\ref{A18})-(\ref{A23}) for (\ref{013}) yields%
\[
-\frac{1}{2(2l+3)}\sqrt{\frac{(l+m+5/2)(l+m+3/2)(l-m+3/2)}{(2l+5)}}\left(
\frac{1}{f_{1}}\right) Y_{l+2,m+1/2}\left( \frac{dG}{dr}-\left( \frac{l+1}{r}%
\right) G\right)
\]%
\[
+\frac{1}{2(2l+3)}\sqrt{\frac{(l-m+3/2)^{2}(l-m+1/2)}{(2l+1)}}\left( \frac{1%
}{f_{1}}\right) Y_{l,m+1/2}\left( \frac{dG}{dr}+\left( \frac{l+2}{r}\right)
G\right)
\]%
\[
+\frac{1}{2(2l+3)}\sqrt{\frac{(l-m+7/2)(l-m+5/2)(l-m+3/2)}{(2l+5)}}\left(
\frac{1}{f_{1}}\right) Y_{l+2,m-3/2}\left( \frac{dG}{dr}-\left( \frac{l+1}{r}%
\right) G\right)
\]%
\[
-\frac{1}{2(2l+3)}\sqrt{\frac{(l+m+1/2)(l+m-1/2)(l-m+3/2)}{(2l+1)}}\left(
\frac{1}{f_{1}}\right) Y_{l,m-3/2}\left( \frac{dG}{dr}+\left( \frac{l+2}{r}%
\right) G\right)
\]%
\[
+\frac{1}{2(2l+3)}\sqrt{\frac{(l-m+1/2)(l-m+3/2)^{2}}{(2l+1)}}\left( \frac{%
R^{\prime }}{2f_{1}}\right) GY_{l,m+1/2}\Delta _{1}
\]%
\[
-\frac{1}{2(2l+3)}\sqrt{\frac{(l+m+5/2)(l+m+3/2)(l-m+3/2)}{(2l+5)}}\left(
\frac{R^{\prime }}{2f_{1}}\right) GY_{l+2,m+1/2}\Delta _{1}
\]%
\[
-\frac{1}{2(2l+3)}\sqrt{\frac{(l+m+1/2)(l+m-1/2)(l-m+3/2)}{(2l+1)}}\left(
\frac{R^{\prime }}{2f_{1}}\right) GY_{l,m-3/2}\Delta _{1}
\]%
\[
+\frac{1}{2(2l+3)}\sqrt{\frac{(l-m+7/2)(l-m+5/2)(l-m+3/2)}{(2l+5)}}\left(
\frac{R^{\prime }}{2f_{1}}\right) GY_{l+2,m-3/2}\Delta _{1}
\]%
\[
-\frac{1}{2(2l+3)}\sqrt{\frac{(l+m+5/2)(l+m+3/2)(l-m+3/2)}{(2l+5)}}\left(
\frac{1}{f_{2}}\right) Y_{l+2,m+1/2}\left( \frac{dG}{dr}-\left( \frac{l+1}{r}%
\right) G\right)
\]%
\[
+\frac{1}{2(2l+3)}\sqrt{\frac{(l-m+3/2)^{2}(l-m+1/2)}{(2l+1)}}\left( \frac{1%
}{f_{2}}\right) Y_{l,m+1/2}\left( \frac{dG}{dr}+\left( \frac{l+2}{r}\right)
G\right)
\]%
\[
-\frac{1}{2(2l+3)}\sqrt{\frac{(l-m+7/2)(l-m+5/2)(l-m+3/2)}{(2l+5)}}\left(
\frac{1}{f_{2}}\right) Y_{l+2,m-3/2}\left( \frac{dG}{dr}-\left( \frac{l+1}{r}%
\right) G\right)
\]%
\[
+\frac{1}{2(2l+3)}\sqrt{\frac{(l+m+1/2)(l+m-1/2)(l-m+3/2)}{(2l+1)}}\left(
\frac{1}{f_{2}}\right) Y_{l,m-3/2}\left( \frac{dG}{dr}+\left( \frac{l+2}{r}%
\right) G\right)
\]%
\[
+\frac{1}{2(2l+3)}\sqrt{\frac{(l-m+1/2)(l-m+3/2)^{2}}{(2l+1)}}\left( \frac{%
R^{\prime }}{2f_{2}}\right) GY_{l,m+1/2}\Delta _{2}
\]%
\[
-\frac{1}{2(2l+3)}\sqrt{\frac{(l+m+5/2)(l+m+3/2)(l-m+3/2)}{(2l+5)}}\left(
\frac{R^{\prime }}{2f_{2}}\right) GY_{l+2,m+1/2}\Delta _{2}
\]%
\[
+\frac{1}{2(2l+3)}\sqrt{\frac{(l+m+1/2)(l+m-1/2)(l-m+3/2)}{(2l+1)}}\left(
\frac{R^{\prime }}{2f_{2}}\right) GY_{l,m-3/2}\Delta _{2}
\]%
\[
-\frac{1}{2(2l+3)}\sqrt{\frac{(l-m+7/2)(l-m+5/2)(l-m+3/2)}{(2l+5)}}\left(
\frac{R^{\prime }}{2f_{2}}\right) GY_{l+2,m-3/2}\Delta _{2}
\]%
\[
+\frac{1}{(2l+3)}\sqrt{\frac{(l+m+5/2)(l+m+3/2)(l-m+3/2)}{(2l+5)}}\left(
\frac{1}{f_{3}}\right) Y_{l+2,m+1/2}\left( \frac{dG}{dr}-\left( \frac{l+1}{r}%
\right) G\right)
\]%
\[
+\frac{1}{(2l+3)}\sqrt{\frac{(l+m+3/2)^{2}(l-m+1/2)}{(2l+1)}}\left( \frac{1}{%
f_{3}}\right) Y_{l,m+1/2}\left( \frac{dG}{dr}+\left( \frac{l+2}{r}\right)
G\right)
\]%
\[
-\frac{1}{(2l+3)}\sqrt{\frac{(l+m+5/2)(l+m+3/2)(l-m+3/2)}{(2l+5)}}\left(
\frac{R^{\prime }}{2f_{3}}\right) GY_{l+2,m+1/2}\Delta _{3}
\]%
\[
-\frac{1}{(2l+3)}\sqrt{\frac{(l+m+3/2)^{2}(l-m+1/2)}{(2l+1)}}\left( \frac{%
R^{\prime }}{2f_{3}}\right) GY_{l,m+1/2}\Delta _{3}
\]%
\begin{equation}
+\left( \frac{E}{f_{4}}-M\right) \sqrt{\frac{(l-m+1/2)}{(2l+1)}}%
FY_{l,m+1/2}=0.  \label{A24}
\end{equation}%
Multiplying both sides by $Y_{l+2,m-3/2},$ $Y_{l,m-3/2},Y_{l+2,m+1/2},$ and $%
Y_{l,m+1/2}$, and performing the integration with respect to the solid angle
gives the four independent equalities%
\begin{equation}
\left( \frac{1}{f_{1}}-\frac{1}{f_{2}}\right) \left[ \frac{dG}{dr}-\left(
\frac{l+1}{r}\right) G\right] +\left[ \frac{(\Delta _{1})}{f_{1}}-\frac{%
(\Delta _{2})}{f_{2}}\right] \frac{R^{\prime }}{2}G=0,  \label{A25}
\end{equation}

\begin{equation}
\left( -\frac{1}{f_{1}}+\frac{1}{f_{2}}\right) \left[ \frac{dG}{dr}+\left(
\frac{l+2}{r}\right) G\right] +\left[ -\frac{(\Delta _{1})}{f_{1}}+\frac{%
(\Delta _{2})}{f_{2}}\right] \frac{R^{\prime }}{2}G=0,  \label{A26}
\end{equation}%
\begin{equation}
\left( \frac{1}{2f_{1}}+\frac{1}{2f_{2}}-\frac{1}{f_{3}}\right) \left[ \frac{%
dG}{dr}-\left( \frac{l+1}{r}\right) G\right] +\left[ \frac{(\Delta _{1})}{%
2f_{1}}+\frac{(\Delta _{2})}{2f_{2}}+\frac{(\Delta _{3})}{f_{3}}\right]
\frac{R^{\prime }}{2}G=0,  \label{A27}
\end{equation}%
and%
\[
\left( \frac{l-m+3/2}{2(2l+3)}\right) \left\{ \left( \frac{1}{f_{1}}+\frac{1%
}{f_{2}}\right) \left[ \frac{dG}{dr}+\left( \frac{l+2}{r}\right) G\right] +%
\left[ \frac{(\Delta _{1})}{f_{1}}+\frac{(\Delta _{2})}{f_{2}}\right] \frac{%
R^{\prime }}{2}G\right\}
\]%
\begin{equation}
+\left( \frac{l+m+3/2}{2l+3}\right) \left\{ \frac{1}{f_{3}}\left[ \frac{dG}{%
dr}+\left( \frac{l+2}{r}\right) G\right] -\frac{(\Delta _{3})}{f_{3}}\frac{%
R^{\prime }}{2}G\right\} +\left( \frac{E}{f_{4}}-M\right) F=0.  \label{A28}
\end{equation}%
These are the equations given in (\ref{023})-(\ref{026}). This completes the
proof of the equivalence.

\section{Proof of Eqs. (\protect\ref{041})-(\protect\ref{044}) and (\protect
\ref{045})-(\protect\ref{048})}

\label{App:AppendixB}

We proceed to prove the equivalence between (\ref{012}) and (\ref{041})-(\ref%
{044}) for the second invariance solution (\ref{017}). Using the relations
in (\ref{A1})-(\ref{A6}), (\ref{012}) can be expressed by%
\[
\frac{1}{2(2l-1)}\sqrt{\frac{(l+m+3/2)(l+m+1/2)(l-m-1/2)}{(2l+1)}}\left(
\frac{1}{f_{1}}\right) Y_{l,m+3/2}\left( \frac{dG}{dr}-\left( \frac{l-1}{r}%
\right) G\right)
\]%
\[
-\frac{1}{2(2l-1)}\sqrt{\frac{(l-m-1/2)(l-m-3/2)(l-m-5/2)}{(2l-3)}}\left(
\frac{1}{f_{1}}\right) Y_{l-2,m+3/2}\left( \frac{dG}{dr}+\left( \frac{l}{r}%
\right) G\right)
\]%
\[
-\frac{1}{2(2l-1)}\sqrt{\frac{(l-m+1/2)(l-m-1/2)^{2}}{(2l+1)}}\left( \frac{1%
}{f_{1}}\right) Y_{l,m-1/2}\left( \frac{dG}{dr}-\left( \frac{l-1}{r}\right)
G\right)
\]%
\[
+\frac{1}{2(2l-1)}\sqrt{\frac{(l+m-3/2)(l+m-1/2)(l-m-1/2)}{(2l-3)}}\left(
\frac{1}{f_{1}}\right) Y_{l-2,m-1/2}\left( \frac{dG}{dr}+\left( \frac{l}{r}%
\right) G\right)
\]%
\[
-\frac{1}{2(2l-1)}\sqrt{\frac{(l-m-1/2)(l-m-3/2)(l-m-5/2)}{(2l-3)}}\left(
\frac{R^{\prime }}{2f_{1}}\right) GY_{l-2,m+3/2}\Delta _{1}
\]%
\[
+\frac{1}{2(2l-1)}\sqrt{\frac{(l+m+3/2)(l+m+1/2)(l-m-1/2)}{(2l+1)}}\left(
\frac{R^{\prime }}{2f_{1}}\right) GY_{l,m+3/2}\Delta _{1}
\]%
\[
+\frac{1}{2(2l-1)}\sqrt{\frac{(l+m-1/2)(l+m-3/2)(l-m-1/2)}{(2l-3)}}\left(
\frac{R^{\prime }}{2f_{1}}\right) GY_{l-2,m-1/2}\Delta _{1}
\]%
\[
-\frac{1}{2(2l-1)}\sqrt{\frac{(l-m+1/2)(l-m-1/2)^{2}}{(2l+1)}}\left( \frac{%
R^{\prime }}{2f_{1}}\right) GY_{l,m-1/2}\Delta _{1}
\]%
\[
-\frac{1}{2(2l-1)}\sqrt{\frac{(l+m+3/2)(l+m+1/2)(l-m-1/2)}{(2l+1)}}\left(
\frac{1}{f_{2}}\right) Y_{l,m+3/2}\left( \frac{dG}{dr}-\left( \frac{l-1}{r}%
\right) G\right)
\]%
\[
+\frac{1}{2(2l-1)}\sqrt{\frac{(l-m-1/2)(l-m-3/2)(l-m-5/2)}{(2l-3)}}\left(
\frac{1}{f_{2}}\right) Y_{l-2,m+3/2}\left( \frac{dG}{dr}+\left( \frac{l}{r}%
\right) G\right)
\]%
\[
-\frac{1}{2(2l-1)}\sqrt{\frac{(l-m+1/2)(l-m-1/2)^{2}}{(2l+1)}}\left( \frac{1%
}{f_{2}}\right) Y_{l,m-1/2}\left( \frac{dG}{dr}-\left( \frac{l-1}{r}\right)
G\right)
\]%
\[
+\frac{1}{2(2l-1)}\sqrt{\frac{(l+m-3/2)(l+m-1/2)(l-m-1/2)}{(2l-3)}}\left(
\frac{1}{f_{2}}\right) Y_{l-2,m-1/2}\left( \frac{dG}{dr}+\left( \frac{l}{r}%
\right) G\right)
\]%
\[
+\frac{1}{2(2l-1)}\sqrt{\frac{(l-m-1/2)(l-m-3/2)(l-m-5/2)}{(2l-3)}}\left(
\frac{R^{\prime }}{2f_{2}}\right) GY_{l-2,m+3/2}\Delta _{2}
\]%
\[
-\frac{1}{2(2l-1)}\sqrt{\frac{(l+m+3/2)(l+m+1/2)(l-m-1/2)}{(2l+1)}}\left(
\frac{R^{\prime }}{2f_{2}}\right) GY_{l,m+3/2}\Delta _{2}
\]%
\[
+\frac{1}{2(2l-1)}\sqrt{\frac{(l+m-1/2)(l+m-3/2)(l-m-1/2)}{(2l-3)}}\left(
\frac{R^{\prime }}{2f_{2}}\right) GY_{l-2,m-1/2}\Delta _{2}
\]%
\[
-\frac{1}{2(2l-1)}\sqrt{\frac{(l-m+1/2)(l-m-1/2)^{2}}{(2l+1)}}\left( \frac{%
R^{\prime }}{2f_{2}}\right) GY_{l,m-1/2}\Delta _{2}
\]

\[
-\frac{1}{(2l-1)}\sqrt{\frac{(l+m-1/2)^{2}(l-m+1/2)}{(2l+1)}}\left( \frac{1}{%
f_{3}}\right) Y_{l,m-1/2}\left( \frac{dG}{dr}-\left( \frac{l-1}{r}\right)
G\right)
\]%
\[
-\frac{1}{(2l-1)}\sqrt{\frac{(l+m-3/2)(l+m-1/2)(l-m-1/2)}{(2l-3)}}\left(
\frac{1}{f_{3}}\right) Y_{l-2,m-1/2}\left( \frac{dG}{dr}+\left( \frac{l}{r}%
\right) G\right)
\]%
\[
+\frac{1}{(2l-1)}\sqrt{\frac{(l+m-1/2)^{2}(l-m+1/2)}{(2l+1)}}\left( \frac{%
R^{\prime }}{2f_{3}}\right) GY_{l,m-1/2}\Delta _{3}
\]%
\[
+\frac{1}{(2l-1)}\sqrt{\frac{(l+m-1/2)(l+m-3/2)(l-m-1/2)}{(2l-3)}}\left(
\frac{R^{\prime }}{2f_{3}}\right) GY_{l-2,m-1/2}\Delta _{3}
\]%
\begin{equation}
-\left( \frac{E}{f_{4}}-M\right) \sqrt{\frac{(l-m+1/2)}{(2l+1)}}%
FY_{l,m-1/2}=0.  \label{B1}
\end{equation}%
Multiplying both sides by $Y_{l,m+3/2},$ $Y_{l-2,m+3/2},Y_{l-2,m-1/2},$ and $%
Y_{l,m-1/2}$, and integrating the equation with respect to the solid angle
gives
\begin{equation}
\left( \frac{1}{f_{1}}-\frac{1}{f_{2}}\right) \left[ \frac{dG}{dr}-\left(
\frac{l-1}{r}\right) G\right] +\left[ \frac{(\Delta _{1})}{f_{1}}-\frac{%
(\Delta _{2})}{f_{2}}\right] \frac{R^{\prime }}{2}G=0,  \label{B2}
\end{equation}

\begin{equation}
\left( -\frac{1}{f_{1}}+\frac{1}{f_{2}}\right) \left[ \frac{dG}{dr}+\frac{l}{%
r}G\right] +\left[ -\frac{(\Delta _{1})}{f_{1}}+\frac{(\Delta _{2})}{f_{2}}%
\right] \frac{R^{\prime }}{2}G=0,  \label{B3}
\end{equation}%
\begin{equation}
\left( \frac{1}{2f_{1}}+\frac{1}{2f_{2}}-\frac{1}{f_{3}}\right) \left[ \frac{%
dG}{dr}+\frac{l}{r}G\right] +\left[ \frac{(\Delta _{1})}{2f_{1}}+\frac{%
(\Delta _{2})}{2f_{2}}+\frac{(\Delta _{3})}{f_{3}}\right] \frac{R^{\prime }}{%
2}G=0,  \label{B4}
\end{equation}%
and%
\[
-\left( \frac{l-m-1/2}{2(2l-1)}\right) \left\{ \left( \frac{1}{f_{1}}+\frac{1%
}{f_{2}}\right) \left[ \frac{dG}{dr}-\left( \frac{l-1}{r}\right) G\right] +%
\left[ \frac{(\Delta _{1})}{f_{1}}+\frac{(\Delta _{2})}{f_{2}}\right] \frac{%
R^{\prime }}{2}G\right\}
\]%
\begin{equation}
-\left( \frac{l+m-1/2}{2l-1}\right) \left\{ \frac{1}{f_{3}}\left[ \frac{dG}{%
dr}-\left( \frac{l-1}{r}\right) G\right] -\frac{(\Delta _{3})}{f_{3}}\frac{%
R^{\prime }}{2}G\right\} -\left( \frac{E}{f_{4}}+M\right) F=0.  \label{B5}
\end{equation}%
These are equations (\ref{041})-(\ref{044}). The proof is completed.

The proof of the equivalence between (\ref{013}) and (\ref{045})-(\ref{048})
is analogous. Using (\ref{A1})-(\ref{A6}), Eq. (\ref{013}) has the
representation%
\[
-\frac{1}{2(2l-1)}\sqrt{\frac{(l+m+1/2)(l+m-1/2)^{2}}{(2l+1)}}\left( \frac{1%
}{f_{1}}\right) Y_{l,m+1/2}\left( \frac{dG}{dr}-\left( \frac{l-1}{r}\right)
G\right)
\]%
\[
+\frac{1}{2(2l-1)}\sqrt{\frac{(l+m-1/2)(l-m-1/2)(l-m-3/2)}{(2l-3)}}\left(
\frac{1}{f_{1}}\right) Y_{l-2,m+1/2}\left( \frac{dG}{dr}+\left( \frac{l}{r}%
\right) G\right)
\]%
\[
+\frac{1}{2(2l-1)}\sqrt{\frac{(l+m-1/2)(l-m+3/2)(l-m+1/2)}{(2l+1)}}\left(
\frac{1}{f_{1}}\right) Y_{l,m-3/2}\left( \frac{dG}{dr}-\left( \frac{l-1}{r}%
\right) G\right)
\]%
\[
-\frac{1}{2(2l-1)}\sqrt{\frac{(l+m-1/2)(l+m-3/2)(l+m-5/2)}{(2l-3)}}\left(
\frac{1}{f_{1}}\right) Y_{l-2,m-3/2}\left( \frac{dG}{dr}+\left( \frac{l}{r}%
\right) G\right)
\]%
\[
+\frac{1}{2(2l-1)}\sqrt{\frac{(l+m-1/2)(l-m-1/2)(l-m-3/2)}{(2l-3)}}\left(
\frac{R^{\prime }}{2f_{1}}\right) GY_{l-2,m+1/2}\Delta _{1}
\]%
\[
-\frac{1}{2(2l-1)}\sqrt{\frac{(l+m+1/2)(l+m-1/2)^{2}}{(2l+1)}}\left( \frac{%
R^{\prime }}{2f_{1}}\right) GY_{l,m+1/2}\Delta _{1}
\]%
\[
-\frac{1}{2(2l-1)}\sqrt{\frac{(l+m-1/2)(l+m-3/2)(l+m-5/2)}{(2l-3)}}\left(
\frac{R^{\prime }}{2f_{1}}\right) GY_{l-2,m-3/2}\Delta _{1}
\]%
\[
+\frac{1}{2(2l-1)}\sqrt{\frac{(l+m-1/2)(l-m+3/2)(l-m+1/2)}{(2l+1)}}\left(
\frac{R^{\prime }}{2f_{1}}\right) GY_{l,m-3/2}\Delta _{1}
\]%
\[
-\frac{1}{2(2l-1)}\sqrt{\frac{(l+m+1/2)(l+m-1/2)^{2}}{(2l+1)}}\left( \frac{1%
}{f_{2}}\right) Y_{l,m+1/2}\left( \frac{dG}{dr}-\left( \frac{l-1}{r}\right)
G\right)
\]%
\[
+\frac{1}{2(2l-1)}\sqrt{\frac{(l+m-1/2)(l-m-1/2)(l-m-3/2)}{(2l-3)}}\left(
\frac{1}{f_{2}}\right) Y_{l-2,m+1/2}\left( \frac{dG}{dr}+\left( \frac{l}{r}%
\right) G\right)
\]%
\[
-\frac{1}{2(2l-1)}\sqrt{\frac{(l+m-1/2)(l-m+3/2)(l-m+1/2)}{(2l+1)}}\left(
\frac{1}{f_{2}}\right) Y_{l,m-3/2}\left( \frac{dG}{dr}-\left( \frac{l-1}{r}%
\right) G\right)
\]%
\[
+\frac{1}{2(2l-1)}\sqrt{\frac{(l+m-1/2)(l+m-3/2)(l+m-5/2)}{(2l-3)}}\left(
\frac{1}{f_{2}}\right) Y_{l-2,m-3/2}\left( \frac{dG}{dr}+\left( \frac{l}{r}%
\right) G\right)
\]%
\[
+\frac{1}{2(2l-1)}\sqrt{\frac{(l+m-1/2)(l-m-1/2)(l-m-3/2)}{(2l-3)}}\left(
\frac{R^{\prime }}{2f_{2}}\right) GY_{l-2,m+1/2}\Delta _{2}
\]%
\[
-\frac{1}{2(2l-1)}\sqrt{\frac{(l+m+1/2)(l+m-1/2)^{2}}{(2l+1)}}\left( \frac{%
R^{\prime }}{2f_{2}}\right) GY_{l,m+1/2}\Delta _{2}
\]%
\[
+\frac{1}{2(2l-1)}\sqrt{\frac{(l+m-1/2)(l+m-3/2)(l+m-5/2)}{(2l-3)}}\left(
\frac{R^{\prime }}{2f_{2}}\right) GY_{l-2,m-3/2}\Delta _{2}
\]%
\[
-\frac{1}{2(2l-1)}\sqrt{\frac{(l+m-1/2)(l-m+3/2)(l-m+1/2)}{(2l+1)}}\left(
\frac{R^{\prime }}{2f_{2}}\right) GY_{l,m-3/2}\Delta _{2}
\]

\[
-\frac{1}{(2l-1)}\sqrt{\frac{(l+m+1/2)(l-m-1/2)^{2}}{(2l+1)}}\left( \frac{1}{%
f_{3}}\right) Y_{l,m+1/2}\left( \frac{dG}{dr}-\left( \frac{l-1}{r}\right)
G\right)
\]%
\[
-\frac{1}{(2l-1)}\sqrt{\frac{(l+m-1/2)(l-m-1/2)(l-m-3/2)}{(2l-3)}}\left(
\frac{1}{f_{3}}\right) Y_{l-2,m+1/2}\left( \frac{dG}{dr}+\left( \frac{l}{r}%
\right) G\right)
\]%
\[
+\frac{1}{(2l-1)}\sqrt{\frac{(l+m+1/2)(l-m-1/2)^{2}}{(2l+1)}}\left( \frac{%
R^{\prime }}{2f_{3}}\right) GY_{l,m+1/2}\Delta _{3}
\]%
\[
+\frac{1}{(2l-1)}\sqrt{\frac{(l+m-1/2)(l-m-1/2)(l-m-3/2)}{(2l-3)}}\left(
\frac{R^{\prime }}{2f_{3}}\right) GY_{l-2,m+1/2}\Delta _{3}
\]%
\begin{equation}
-\left( \frac{E}{f_{4}}-M\right) \sqrt{\frac{(l+m+1/2)}{(2l+1)}}%
FY_{l,m+1/2}=0.  \label{B6}
\end{equation}%
Multiplying both sides by $Y_{l,m-3/2},$ $Y_{l-2,m-3/2},Y_{l-2,m+1/2},$ and $%
Y_{l,m+1/2}$, and integrating both sides of the equation with respect to the
solid angle gives%
\begin{equation}
\left( \frac{1}{f_{1}}-\frac{1}{f_{2}}\right) \left[ \frac{dG}{dr}-\left(
\frac{l-1}{r}\right) G\right] +\left[ \frac{(\Delta _{1})}{f_{1}}-\frac{%
(\Delta _{2})}{f_{2}}\right] \frac{R^{\prime }}{2}G=0,  \label{B7}
\end{equation}

\begin{mathletters}
\begin{equation}
\left( -\frac{1}{f_{1}}+\frac{1}{f_{2}}\right) \left[ \frac{dG}{dr}+\frac{l}{%
r}G\right] +\left[ -\frac{(\Delta _{1})}{f_{1}}+\frac{(\Delta _{2})}{f_{2}}%
\right] \frac{R^{\prime }}{2}G=0,  \label{B8}
\end{equation}%
\end{mathletters}
\begin{equation}
\left( \frac{1}{2f_{1}}+\frac{1}{2f_{2}}-\frac{1}{f_{3}}\right) \left[ \frac{%
dG}{dr}+\frac{l}{r}G\right] +\left[ \frac{(\Delta _{1})}{2f_{1}}+\frac{%
(\Delta _{2})}{2f_{2}}+\frac{(\Delta _{3})}{f_{3}}\right] \frac{R^{\prime }}{%
2}G=0,  \label{B9}
\end{equation}%
and%
\[
-\left( \frac{l+m-1/2}{2(2l-1)}\right) \left\{ \left( \frac{1}{f_{1}}+\frac{1%
}{f_{2}}\right) \left[ \frac{dG}{dr}-\left( \frac{l-1}{r}\right) G\right] +%
\left[ \frac{(\Delta _{1})}{f_{1}}+\frac{(\Delta _{2})}{f_{2}}\right] \frac{%
R^{\prime }}{2}G\right\}
\]%
\begin{equation}
-\left( \frac{l-m-1/2}{2l-1}\right) \left\{ \frac{1}{f_{3}}\left[ \frac{dG}{%
dr}-\left( \frac{l-1}{r}\right) G\right] -\frac{(\Delta _{3})}{f_{3}}\frac{%
R^{\prime }}{2}G\right\} -\left( \frac{E}{f_{4}}-M\right) F=0.  \label{B10}
\end{equation}%
These are Eqs. (\ref{045})-(\ref{048}). Hence the equivalence is proved.
\newline
\newline
{\centerline{ACKNOWLEDGMENTS}}
\center {The author would like to thank
C. R. Harrington for her careful reading of the manuscript.
My work has been supported by the National
Science Council of Taiwan under contract No. NSC 102-2112-M-110-012-MY3.} 

\newpage
\begin{figure}[hbt]
\includegraphics[width=5in]{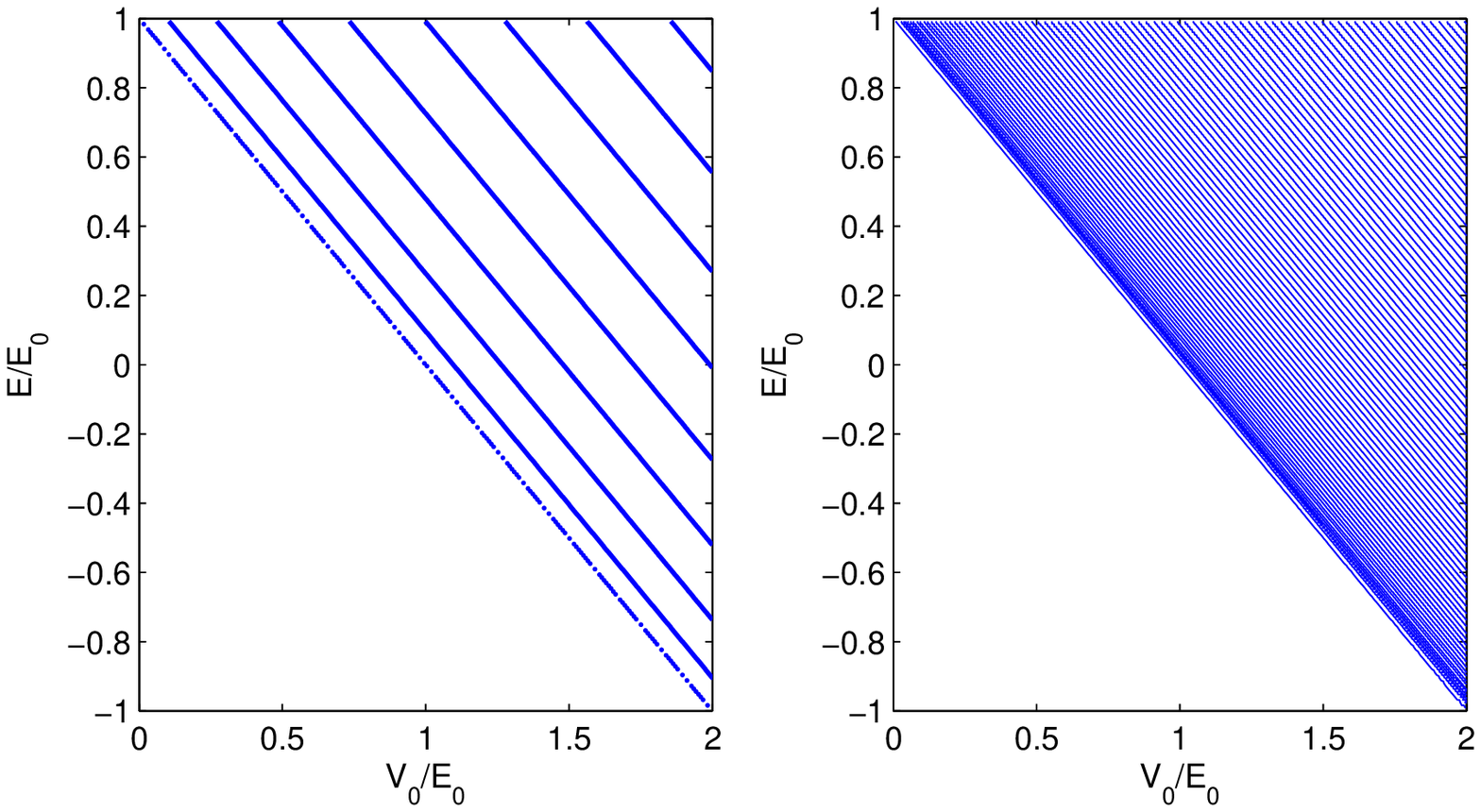}
\caption{(Color online) Energy levels of trapped fermion states in the spacetime cage. Left: The figure shows the energy levels of the $s$ states as a function of the potential strength which is determined by the proper time, see (\ref{0104a}), where $E_0=Mc^2$, the rest energy of a trapped spin-1/2 fermion. The radius of the cage is $r_b=10\lambda_F$, where $\lambda_F=\hbar/Mc$ is the Compton wavelength of the fermions. The deeper the cage, the more bound levels it will allow. Right: The pattern shows the energy spectrum of the cage with radius $r_b=100\lambda_F$.}
\label{fig1}
\end{figure}
\begin{figure}[hbt]
\includegraphics[width=5in]{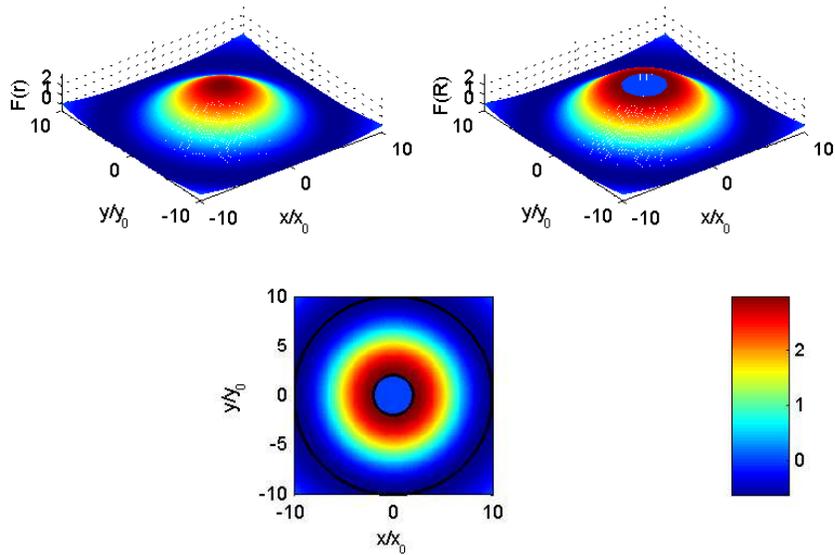}
\caption{(Color online)
A bound state accompanying an invisible region.
Since the radial amplitude $F$ is spherically symmetric, it suffices to show its distribution at the $x$-$y$ plane.
Top left: The component $F(r)$ for $1s_{1/2}$ state in a regular spherical well, where $x_0=y_0=\lambda_F$. Top right: The presented cage creates a zero distribution region of the component around the center of the bound state, where the radius of the cage is $10\lambda_F$, and the radius of the zero amplitude region is chosen as $2\lambda_F$. To plot the patterns, the depth of the cage and the bound energy are chosen as $V_0/E_0=0.1043$, and $E/E_0=0.992$ which are extracted from the numerical data in Fig. 1. The amplitude is not normalized to clearly exhibit the solid profiles. The bottom pattern shows the projection of the top right on the $x$-$y$ plane.}
\label{fig2}
\end{figure}
\begin{figure}[hbt]
\includegraphics[width=5in]{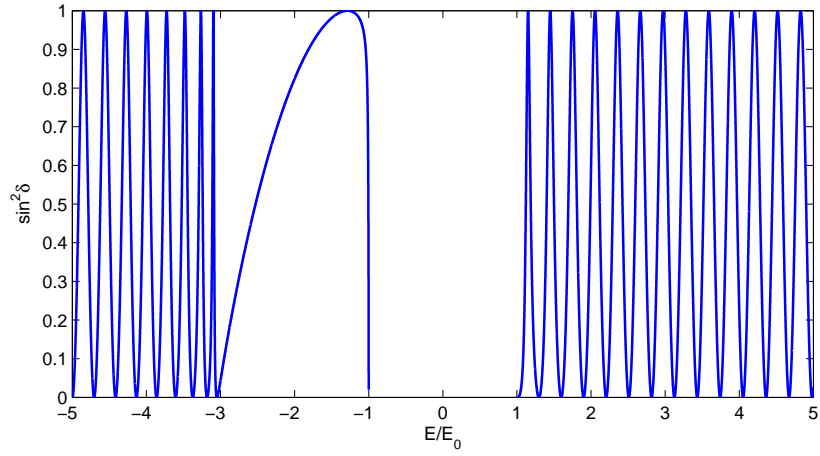}
\caption{(Color online)
The phase shift of $s$ states for the fermionic cage. The zeros of $\delta$ correspond to the resonances, where one has a larger probability of finding the fermions in the cage.}
\label{fig3}
\end{figure}
\end{document}